\documentclass[useAMS,usenatbib]{mn2e}
\pdfoutput=1
\usepackage[english]{babel}
\usepackage{amsfonts,amsmath,amssymb}
\usepackage{times}
\usepackage{paralist}
\usepackage{graphicx}
\usepackage{color}
\usepackage{tikz}
\usepackage[colorlinks,linkcolor=blue,citecolor=blue,urlcolor=blue,pdftitle={Means of confusion: how pixel noise affects shear estimates for weak gravitational lensing},pdfauthor={Melchior \& Viola}]{hyperref}
\title[How pixel noise affects shear estimates for weak gravitational lensing]{Means of confusion: how pixel noise affects shear estimates for weak gravitational lensing}
\author[Melchior \& Viola]{P. Melchior$^{1,2}$\thanks{E-mail: melchior.12@osu.edu}, M. Viola$^{3}$\\
$^{1}$Center for Cosmology and Astro-Particle Physics, The Ohio State University, 191 W. Woodruff Ave., Columbus, Ohio 43210, USA\\
$^{2}$Department of Physics, The Ohio State University, 191 W. Woodruff Ave., Columbus, Ohio 43210, USA\\
$^{3}$Institute for Astronomy,  University of Edinburgh, Royal Observatory, Blackford Hill, Edinburgh, EH9 3HJ, United Kingdom\\
}

\begin{document}

\date{Accepted 2012 May 23. Received 2012 May 18; in original form 2012 April 23}

\pagerange{\pageref{firstpage}--\pageref{lastpage}} \pubyear{2012}

\maketitle
\label{firstpage}

\begin{abstract}
Weak-lensing shear estimates show a troublesome dependence on the apparent brightness of the galaxies used to measure the ellipticity: In several studies, the amplitude of the inferred shear falls sharply with decreasing source significance. This dependence limits the overall ability of upcoming large weak-lensing surveys to constrain cosmological parameters.\\
We seek to provide a concise overview of the impact  of pixel noise on weak-lensing measurements, covering the entire path from noisy images to shear estimates. We show that there are at least three distinct layers, where pixel noise not only obscures but biases the outcome of the measurements: 1) the propagation of pixel noise to the non-linear observable ellipticity; 2) the response of the shape-measurement methods to limited amount of information extractable from noisy images; and 3) the reaction of shear estimation statistics to the presence of noise and outliers in the measured ellipticities.\\
We identify and discuss several fundamental problems and show that each of them is able to introduce \emph{biases in the range of a few tenths to a few percent} for galaxies with typical significance levels. Furthermore, all of these biases do not only depend on the brightness of galaxies but also on their ellipticity, with more elliptical galaxies often being harder to measure correctly. We also discuss existing possibilities to mitigate and novel ideas to avoid the biases induced by pixel noise. We present a new shear estimator that shows a more robust performance for noisy ellipticity samples. Finally, we release the open-source {\sc python} code to predict and efficiently sample from the noisy ellipticity distribution and the shear estimators used in this work at \href{https://github.com/pmelchior/epsnoise}{this URL}.
\end{abstract}


\section{Introduction}
Current and in particular upcoming wide-field imaging surveys such as the Dark Energy Survey\footnote{\url{http://www.darkenergysurvey.org/}}, the KIlo Degree Survey\footnote{\url{http://kids.strw.leidenuniv.nl/}}, the Hyper Suprime Camera Survey, Euclid\footnote{\url{http://sci.esa.int/euclid}}, and the Wide-Field Infrared Survey Telescope\footnote{\url{http://wfirst.gsfc.nasa.gov/}} require highly accurate shape measurement methods to reach the forecasted accuracy of cosmological parameters constraints, e.g. for the energy density of matter $\Omega_m$, the normalization of the matter power spectrum $\sigma_8$, the Dark Energy equation-of-state parameter $w$ and its variation with time. The currently most demanding lensing method, the cosmic shear two-point correlation function, allows for multiplicative errors (defined as the deviation of actual shear from the measurement by a factor $m$), with $|m|$ not larger than some per mille \citep[e.g][]{Huterer06.1,Amara08.1}. But with increasing survey volumes also traditionally less demanding techniques, such as stacked cluster lensing, will require $|m|$ of order 1\% \citep{Weinberg12.1}. Shear measurement methods known so far can reach these requirement in certain regimes, for instance for well-resolved and bright galaxies. However, they commonly struggle with small and especially with faint galaxies \citep{Massey07.1,Bridle10.1, Kitching12.1}. Many suggestions have been brought forward as to why prominent pixel noise hampers shape measurements, but often it was difficult to disentangle the causes from their consequences.

\citet{Bernstein02.1} computed that this so-called \emph{noise rectification bias} generally scales inversely with the second power of the object's significance. \citet{Hirata04.1} obtained an analytic description of the dependence of this bias on the sizes of galaxies, but this derivation only applies to their adaptive moment-based measurement method. Recently, \citet{Refregier12.1} showed that biases in the maximum-likelihood estimators of model-fitting approaches are a direct consequence of the presence of non-linear fit parameters. They also provide an analytic expression for the bias of the ellipticity estimate, valid for Gaussian-shaped galaxies and Point-spread functions.

We seek to generalize and extend these previous findings. In particular, we choose an approach, which is as method-independent as possible, offering insights in the several ways in which pixel noise obscures the estimation of weak gravitational shear.

\subsection*{Approach}
Throughout this work, we aim to identify conceptual problems $\mathcal{P}$ for the shape estimation task, which give rise to deviations of the measured source ellipticity $\bepsilon^\prime(\mathcal{P})$ from the true ellipticity $\bepsilon$. We assume a probabilistic approach, where we inspect the probability distribution of $\mathcal{P}$ under noise, $p(\mathcal{P}|\nu)$, with $\nu$ denoting some suitable characterization of the significance of the measurement, and its consequence, the probability distribution of the measured ellipticity caused by $\mathcal{P}$,
\begin{equation}
p_\mathcal{P}(\bepsilon^\prime| \nu) = \int d\mathcal{P}\ \bepsilon^\prime(\mathcal{P})\, p(\mathcal{P}|\nu).
\end{equation}
The ellipticity distribution is thus given by the impact $\mathcal{P}$ has on $\bepsilon^\prime$, weighted by the probability that $\mathcal{P}$ actually occurs in a measurement with significance $\nu$. The key assumption here is that we could measure the function $\bepsilon^\prime(\mathcal{P})$ perfectly, i.e. without pixel noise, while the action of the noise is entirely contained in the width and shape of the probability distribution of $\mathcal{P}$.

In practice, both $\bepsilon^\prime(\mathcal{P})$ and $p(\mathcal{P}|\nu)$ additionally depend on the apparent shape of the source, i.e. its intrinsic shape and the effects of the convolution with the Point-spread function. We therefore introduce a parameterization of the apparent source morphology with a parameter vector $\vec{\theta}$. 

Although the true ellipticity $\bepsilon$ could be regarded as one of these source parameters, we choose to make the dependency of $\bepsilon^\prime$ on $\bepsilon$ explicit, such that our most general form of the ellipticity distribution caused by $\mathcal{P}$ reads as
\begin{equation}
\label{eq:eps_distr}
p_\mathcal{P}(\bepsilon^\prime| \bepsilon, \vec{\theta}, \nu) = \int d\mathcal{P}\ \bepsilon^\prime(\mathcal{P} | \bepsilon, \vec{\theta})\, p(\mathcal{P}| \bepsilon, \vec{\theta}, \nu).
\end{equation}
We can now define when we consider an ellipticity measurement to be \emph{unbiased} by $\mathcal{P}$, namely if
\begin{equation}
\label{eq:biasdef}
\langle p_\mathcal{P}(\bepsilon^\prime | \bepsilon, \vec\theta, \nu)\rangle = \bepsilon,
\end{equation}
where the average is taken over independent noise realizations of images of identical galaxies, parameterized by $(\bepsilon, \vec\theta)$. In \autoref{sec:error_propagation} and \autoref{sec:measurements} we are going to inspect cases, where, for different reasons, biases occur for fixed $\bepsilon$, whereas in \autoref{sec:statistics} we are going to discuss the application of statistics to samples of noisy ellipticity estimates, whose true values are drawn from an underlying distribution $p(\bepsilon)$. We conclude in \autoref{sec:discussion}.

\section{Non-linear error propagation}
\label{sec:error_propagation}

An object's ellipticity is necessarily a non-linear quantity since any definition needs to invoke a ratio of the two key parameters of the geometric ellipse, its semi-major and semi-minor axes. As with any parameter that depends non-linearly on the data, even a symmetric distribution of noise for each data points translates into much more complicated, in general asymmetric and skewed, distribution of the ellipticity. \citet{Refregier12.1} describe this generic problem specifically for the case of model-dependent galaxy shape measurements and work out the bias on size and ellipticity of the best-fit model that occurs even if the functional form of the galactic shape is perfectly known. We extend their finding by providing a theoretical form of the distribution of noisy ellipticity estimates that works for model- and moment-based approaches. The key ingredient of the derivation is the understanding that any attempt to measure the ellipticity is affected by the spurious ellipticity pattern of the noise realization recorded in the image, which is entirely describable by its moments, whereas spatial models for arbitrary noise configurations are not meaningful.

The second-order moments of the light distribution $I(\mathbf{x})$, centered at $\bar{\mathbf{x}}$,
\begin{equation}
\label{eq:2ndmoments}
Q_{ij} \equiv \int d^2 x\ I(\mathbf{x}) \bigl(x_i - \bar{x}_i\bigr)\,\bigl(x_j -\bar{x}_j\bigr),
\end{equation}
give rise to a number of ellipticity estimators, two of which are in widespread use,
\begin{subequations}
\label{eq:ellipticities}
\begin{align}
\label{eq:chi}
\bchi &\equiv \frac{Q_{11} - Q_{22} + 2\mathrm{i}Q_{12}}{Q_{11} + Q_{22}} \ \ \ \text{and}\\
 \label{eq:ellipticity}
\bepsilon &\equiv \frac{Q_{11}-Q_{22}+2iQ_{12}}
   {Q_{11}+Q_{22}+2\sqrt{Q_{11}Q_{22}-Q_{12}^2}}.
\end{align}
\end{subequations}
We review their advantages and disadvantages in \autoref{sec:eps_chi}. In summary, $\bepsilon$ is, at least theoretically, an unbiased estimator of the shear, while $\bchi$ has a much more favorable distribution under noise, but requires higher-order corrections when used as shear estimator. Since we seek to obtain a distribution of ellipticities under noise, we choose $\bchi$ for the derivation, but continue to use $\bepsilon$ when referring to the proper ellipticity of an object.

In presence of noise \autoref{eq:2ndmoments} needs to be modified,
\begin{equation}
\label{eq:weighted_moments}
Q_{ij}^\prime=\int d^2 x\ W(\mathbf{x}-\bar{\mathbf{x}})\bigl[I(\mathbf{x})+n(\mathbf{x})\bigr] \bigl(x_i - \bar{x}_i\bigr)\,\bigl(x_j -\bar{x}_j\bigr)
\end{equation}
where $W$ is a centered weight function and $n$ is the (uncorrelated) noise term, which, in the background-dominated limit of faint objects, can be assumed to be drawn from an uncorrelated Gaussian distribution with variance $\sigma^{2}_{n}$,
\begin{equation}
\label{eq:noise}
n \sim \mathcal{N}(0,\sigma_n^2),\ \ \langle n(\mathbf{x}) n(\mathbf{x}^\prime)\rangle = \sigma_n^2 \delta(\mathbf{x}-\mathbf{x}^\prime).
\end{equation}
Even though weight functions are only explicitly present in moment-based approaches, model-based approaches effectively weigh pixels according to their distance and the shape of the employed model, so they too make use of a weight function.

Since moments are linear in the image data, they inherit the Gaussian error distribution from $n$. By the same token, the algebraic sum of moments defining the numerator and the denominator in \autoref{eq:chi} are Gaussian distributed. But what about their ratio? 
It is basic knowledge in statistics that the ratio $t$ of two uncorrelated Gaussian variates with mean of zero and unit variance, $\mathcal{N}(0,1)$, is distributed according to the Cauchy distribution, 
\begin{equation}
\label{eq:cauchy}
\mathcal{C}(t) = \frac{1}{\mathrm{\pi}(1+t^2)}.
\end{equation}
But for the ellipticity-measurement problem, the combination of moments defining $\bchi$ do not have zero mean nor are they uncorrelated. It is obvious that at least the denominator of $\bchi$ does not vanish for a source with non-negative brightness distribution. Also, the origin of the correlation quickly becomes apparent, when we choose a frame such that $|\bchi| = \chi_1 > 0$, which can always be realized by a suitable rotation. We can write the mapping of $(Q_{11}, Q_{22})$ onto new variables $(w,z)$, the denominator and numerator of $\chi_1$, as a linear operator
\begin{equation}
\mathsf{M} = \begin{pmatrix} 1& 1\\1&-1\end{pmatrix}.
\end{equation}
The covariance matrix of $(w,z)$ is then given by
\begin{equation}
\label{eq:Swz0}
\begin{split}
\mathsf{S}_{w,z} & = \mathsf{M} \mathsf{S}_{11,22} \mathsf{M^T} = \begin{pmatrix} \sigma_{11}^2 + \sigma_{22}^2 & \sigma_{11}^2 - \sigma_{22}^2 \\ \sigma_{11}^2 - \sigma_{22}^2 & \sigma_{11}^2 + \sigma_{22}^2\end{pmatrix}\\
&\equiv \begin{pmatrix} \sigma_w^2 & \rho\sigma_w\sigma_z \\ \rho\sigma_w\sigma_z & \sigma_z^2\end{pmatrix},
\end{split}
\end{equation}
where $\mathsf{S}_{11,22}=\textrm{Diag}(\sigma_{11}^2, \sigma_{22}^2)$ denotes the (diagonal) covariance matrix of the moments $Q_{ii}$.
Thus, the correlation between $w$ and $z$,
\begin{equation}
\label{eq:fit}
\rho = \frac{\sigma_{11}^2 - \sigma_{22}^2}{\sigma_w\sigma_z},
\end{equation}
only vanishes if the variances of the two moments $Q_{11}$ and $Q_{22}$ are equal, i.e. for circular galaxies. However, if the weight function $W$ is adjusted to match the apparent shape of the galaxy, then the variances will generally be different. In our case with $\chi_1>0$, the pixels along the 1-direction (the semi-major axis) have a larger weight than those perpendicular to it. It follows that $\sigma_{11} > \sigma_{22}$. Thus, the variance of both $w$ and $z$ are driven by $\sigma_{11}$, and they become more and more correlated the larger $\chi_1$ gets.

\begin{figure}
\includegraphics[width=\linewidth]{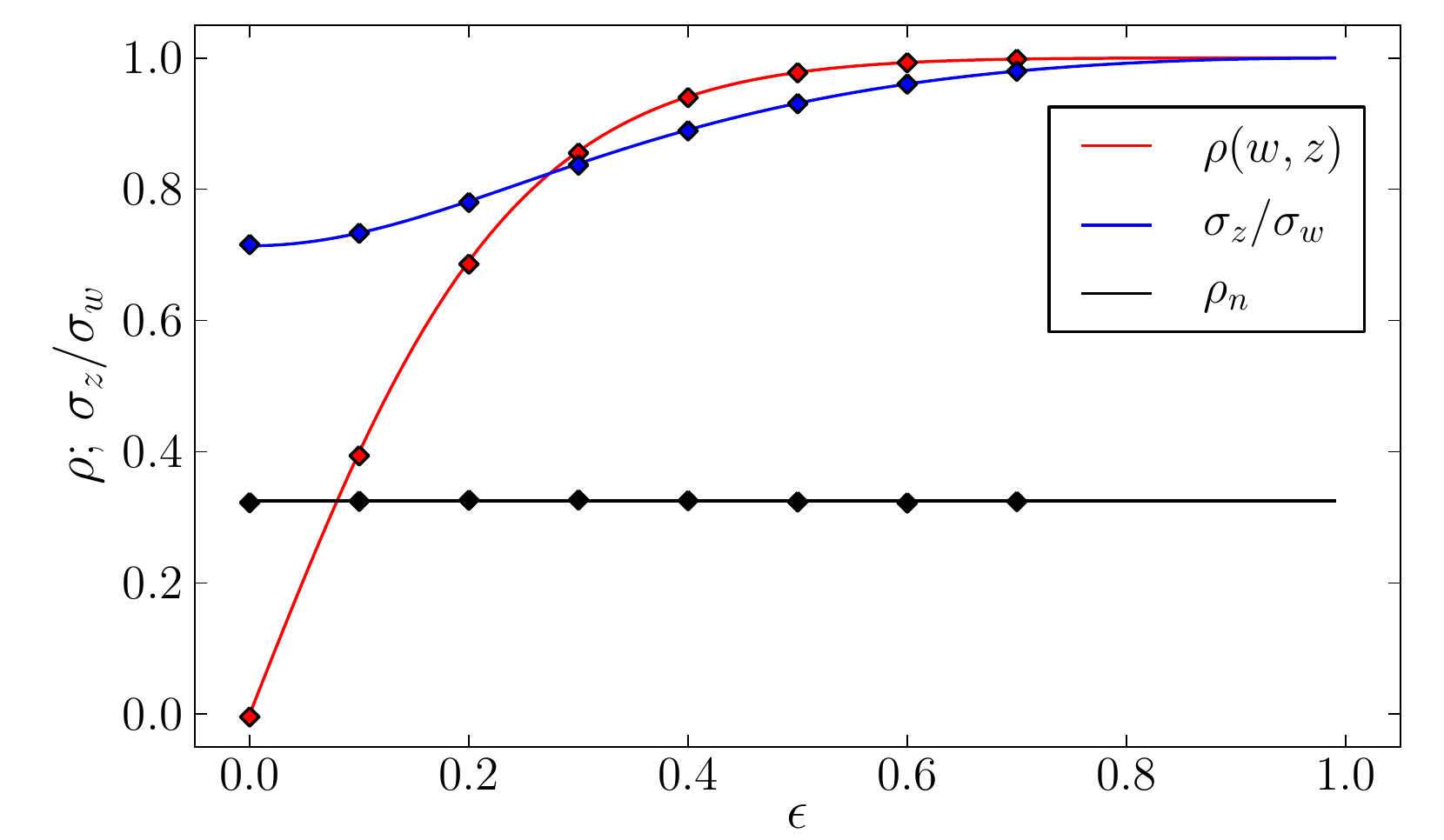}
\caption{Correlation coefficient $\rho_n$ between noisy moments $Q_{11}^\prime$ and $Q_{22}^\prime$ (black line) and between the moment combinations $z=Q^\prime_{11}-Q^\prime_{22}$ and $w=Q^\prime_{11}+Q^\prime_{22}$ (red line) as a function of the ellipticity of the Gaussian weighting function. Due to $\rho_n\approx 0.325$, the errors of $w$ and $z$ are different (their ratio: blue line). Dots indicate the measurement of these quantities in numerical tests with 10,000 noise realizations for each galaxy image. Variation of the size and radial profile of the weight function amounted to sub-percent changes on the quantities shown above.}
\label{fig:correlation}
\end{figure}

But this picture is not yet entirely complete because for \autoref{eq:Swz0} we assumed the moments $Q_{11}$ and $Q_{22}$ to be uncorrelated, but for any given image they are determined by the same noise realization. Imagine a perfectly noise-free image, where we add a positive noise fluctuation in only one pixel. According to \autoref{eq:weighted_moments}, both $Q_{11}^\prime$ and $Q_{22}^\prime$ will then be larger than their noise-free counterparts since their distance-weighting factors $(x_i - \bar{x}_i)^2$ are positive or zero. Effectively, the errors of these two moments are not independent. As we show in \autoref{sec:sampling} in the case of elliptical objects with Gaussian radial profiles of width $s$, the covariance matrix of $Q_{11}$ and $Q_{22}$ is given by
\begin{equation}
\mathsf{S}_{11,22} = \begin{pmatrix}
\sigma_{11}^2 & \sigma_{12}^2\\
\sigma_{12}^2 & \sigma_{22}^2
\end{pmatrix} \propto 
\begin{pmatrix} 
\frac{3 s^6}{(1-\epsilon)^5(1+\epsilon)} & \frac{1 s^6}{(1-\epsilon)^3(1+\epsilon)^3}\\
\frac{1 s^6}{(1-\epsilon)^3(1+\epsilon)^3} & \frac{3 s^6}{(1-\epsilon)(1+\epsilon)^5}\\
\end{pmatrix},
\end{equation}
which automatically implies that the correlation coefficient between $Q_{11}$ and $Q_{22}$,
\begin{equation}
\label{eq:rho_13}
\rho_n = \frac{\sigma_{12}^2}{\sigma_{11}\sigma_{22}} = \frac{1}{3},
\end{equation}  
independent of size and ellipticity of the object. Numerical tests showed that this correlation is largely unaffected even by changes to the radial profile of the weight function (see \autoref{fig:correlation}). The consequence of this correlation is a modification of \autoref{eq:Swz0},
\begin{equation}
\label{eq:Swz}
\mathsf{S}_{w,z}  = \begin{pmatrix} \sigma_{11}^2 + \sigma_{22}^2 + 2\rho_n\sigma_{11}\sigma_{22}& \sigma_{11}^2 - \sigma_{22}^2 \\ \sigma_{11}^2 - \sigma_{22}^2 & \sigma_{11}^2 + \sigma_{22}^2  - 2\rho_n\sigma_{11}\sigma_{22}\end{pmatrix},
\end{equation}
which does not alter $\rho$ but the variances of $w$ and $z$. The ratio $\sigma_z/\sigma_w$ is also shown in \autoref{fig:correlation}, together with $\rho$. It is remarkable how strongly correlated the two moment combinations become even at modest ellipticities. Again, this result shows only marginal changes under variation of weighting function size or radial profile.

\begin{figure}
\includegraphics[width=\linewidth]{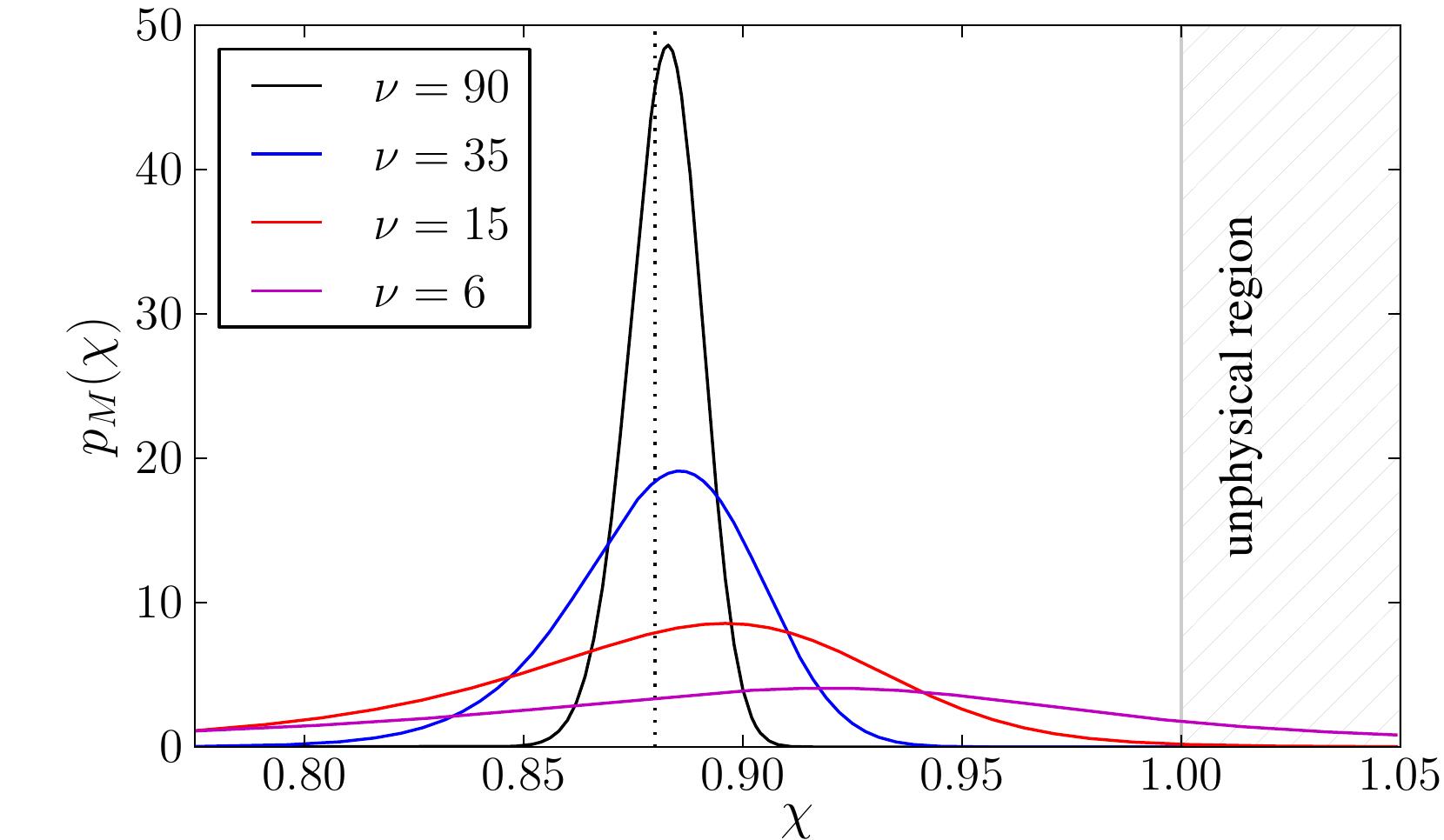}
\caption{Marsaglia distribution (\autoref{eq:marsaglia}) of the ellipticity $\chi$ of a galaxy with true ellipticity $\chi=0.88$ (equivalent to $\epsilon=0.6$, indicated by the vertical dotted line) as a function of image significance $\nu$. The correlation strength $\rho=0.993$ is characteristic of this value of $\epsilon$ (cf. \autoref{eq:fit} and \autoref{fig:correlation}). The hatched region indicates the unphysical but possible range of outliers with $|\chi|\geq1$.}
\label{fig:Marsaglia}
\end{figure}

We are now equipped with variances and correlation of $w$ and $z$ and want to know the distribution of the ratio $t=\tfrac{z}{w}$. \citet{Marsaglia65.1,Marsaglia06.1} proved that the ratio of correlated Gaussian variates $w\sim\mathcal{N}(\mu_w,\sigma_w)$ and $z~\sim\mathcal{N}(\mu_z,\sigma_z)$ with correlation coefficient $\rho$ is given by
\begin{equation}
\label{eq:marsaglia}
p_M(t) = r\, f\bigl(r(s-t)\bigr),
\end{equation}
with constants defined as
\begin{equation}
r=\frac{\sigma_{w}}{\sigma_{z}\sqrt{1-\rho^2}} \ \ \text{and}\ \ s=\rho\frac{\sigma_{z}}{\sigma_{w}}. 
\end{equation}
The function $f(\tau)$ describes the probability distribution of $\frac{a+x}{b+y}$, with $x, y \sim \mathcal{N}(0,1)$:
\begin{equation}
\label{eq:f_tau}
f(\tau)=\frac{\mathrm{e}^{-\frac{1}{2}(a^2+b^2)}}{\pi(1+\tau^2)}\Bigg(1+\frac{\pi}{2}q\,\mathrm{e}^{\frac{1}{2}q^2}\mathrm{Erf}\Bigg(\sqrt{\frac{1}{2}}q\Bigg)\Bigg),
\end{equation}
where
\begin{equation}
q=\frac{b+a\tau}{\sqrt{1+\tau^2}},\ a=\frac{\mu_{z}/\sigma_{z}-\rho \mu_{w}/\sigma_{w}}{\sqrt{1-\rho^2}}, \ \text{and}\ \ b=\frac{\mu_{w}}{\sigma_{w}}.
\end{equation}
We will refer to \autoref{eq:marsaglia} as the \textit{Marsaglia distribution}.\footnote{Unsurprisingly, in the case of uncorrelated variates $w, z  \sim \mathcal{N}(0,1)$, the first term of \autoref{eq:f_tau} recovers the Cauchy distribution of \autoref{eq:cauchy}.} For the ellipticity distribution we only have to substitute 
\begin{equation}
\mu_w = Q_{11}+Q_{22},\ \mu_z=Q_{11}-Q_{22}
\end{equation}
and to take variances and correlation from \autoref{eq:Swz}.
 
A non-vanishing correlation $\rho> 0$ has important consequences, foremost $s>0$ and thus a shift of the peak of the ellipticity distribution towards higher values of $\chi$. In \autoref{fig:Marsaglia}, we show the distribution for a fixed ellipticity, i.e. fixed $\rho$, as a function of the image significance $\nu$, varied in steps of one magnitude.\footnote{For the entire paper, we use the definition of $\nu$ from \citet{Erben01.1}.} We can see that even for fairly bright galaxies, a substantial shift occurs, which grows with decreasing $\nu$. Less elliptical galaxies exhibit a weaker but still noticeable shift of the peak. 

\begin{figure}
\includegraphics[width=\linewidth]{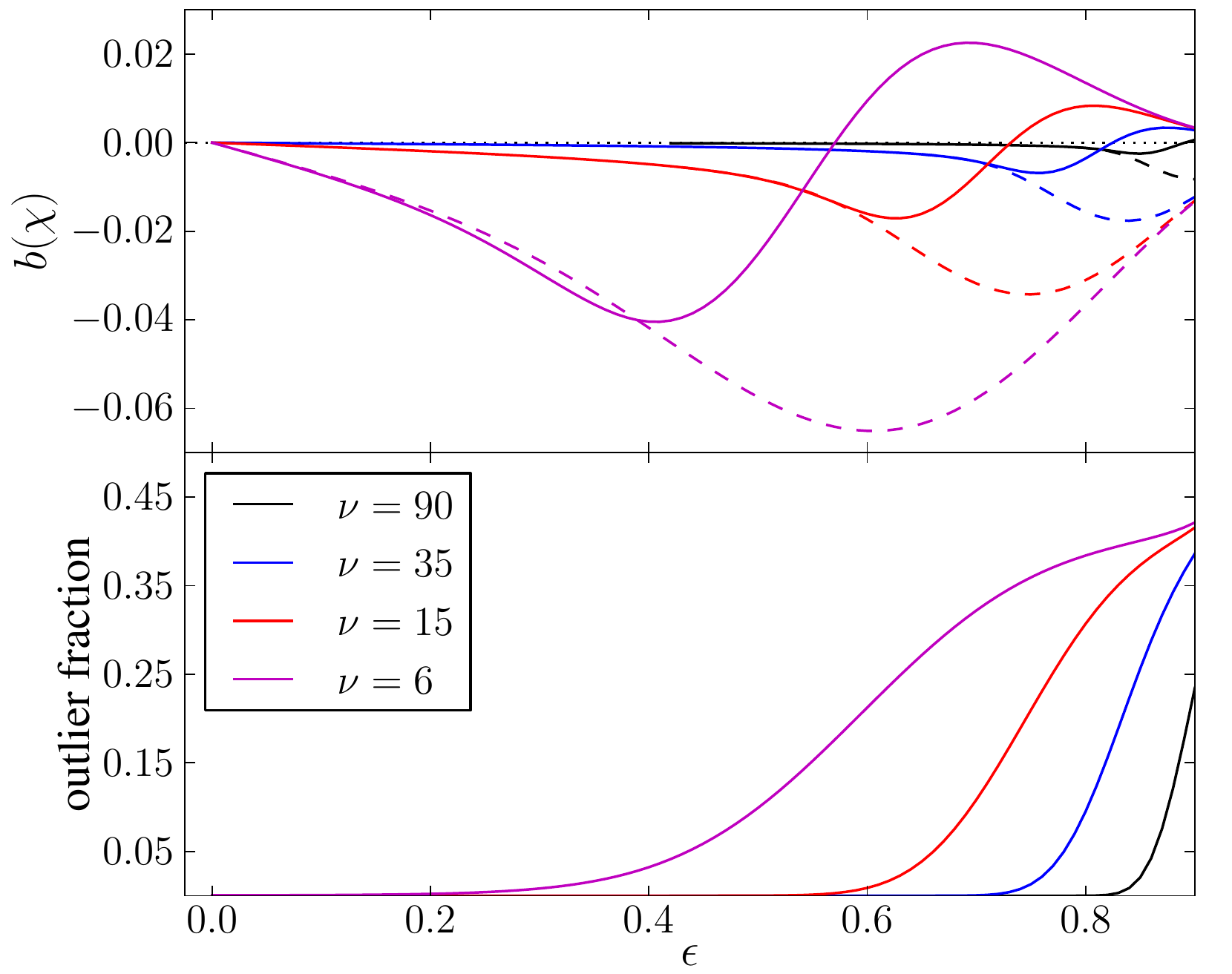}
\caption{\emph{Top:} Ellipticity bias $b(\chi)$, defined as the difference between the mean of the Marsaglia distribution of \autoref{eq:marsaglia} and the ellipticity $\chi$, as a function of the proper ellipticity $\epsilon$ for different significance levels (solid lines). The restriction of the integral to within $|\chi|=1$ leads to stronger biases (dashed lines). \emph{Bottom:} The fraction of measurements with $|\chi^\prime|>1$.}
\label{fig:MarsagliaBias}
\end{figure}

There are two other remarkable features of the Marsaglia distribution. First, it is not only shifted but in general also skewed (this is true even for $\rho=0$) with a long tail towards lower values of $|\chi|$. It is thus not obvious whether the expectation value $\langle p_M(\bchi^\prime | \bepsilon, \vec\theta, \nu)\rangle$ deviates from $\bchi$, constituting a bias according to \autoref{eq:biasdef}. The top panel of \autoref{fig:MarsagliaBias} shows the integral over the entire distribution minus the true value of $\chi$ as a function of $\epsilon$ and different values of $\nu$ (solid lines). As expected, there is almost no bias for bright objects (black solid line), even up to large ellipticities. But with increasing noise level, the bias becomes initially negative (long tail dominates) before it turns positive (shift of the peak dominates). That means, the simple fact that the ellipticity is related to the image data in a non-linear way constitutes a bias of remarkable amplitude and non-trivial behavior.

Similar findings, namely a shift of the peak and a skewed distribution, are reported by \citet{Refregier12.1} and \citet{Kacprzak12.1}, but our derivation does not require us to adopting a model-fitting approach or Gaussian likelihoods. It is thus not restricted to a particular shape of the galaxy and can also deal with convolutions with arbitrary PSF shapes as long as the process can be described in terms of moments \citep[cf.][]{Melchior11.1}. It is, however, not entirely obvious how to quantitatively compare to their distribution of maximum-likelihood estimators of $\bepsilon$.

Second, the Marsaglia distribution it is not bound by $|\chi| \leq 1$. In fact, it inherits the wide Cauchy-type wings and is thus capable of generating outliers with unbounded errors. Still, most errors are comparatively small such that outliers most often originate from galaxies with large initial ellipticities, rendering the outlier fraction strongly ellipticity dependent (bottom panel \autoref{fig:MarsagliaBias}). With increasing noise level, ever smaller ellipticities become possible outliers. The exclusion of outliers from the integral over the distribution $p_M(\chi)$ leads to much stronger and mostly negative biases (dashed lines in the top panel) as the positive impact of the shifted peak becomes limited. How measurement codes and shear statistics respond to such outliers will be a reoccurring topic in the remainder of this work. 

\section{Measurement-related ellipticity biases}
\label{sec:measurements}

As we laid out above, an ellipticity measurement cannot provide an unbiased result, simply because the error distribution is shifted and skewed for any ellipticity $|\bepsilon| > 0$. Instead of requiring a shape measurement method to yield unbiased ellipticity estimates, we should rather require that it reproduces the theoretically expected noisy distribution. In this section, we argue that in general not even that is possible because the attempt to measure a shape necessarily contributes its own uncertainties and, for non-linear shape parameters such as the galaxy size, its own biases \citep{Refregier12.1}.

Even though the biases arising from the limited amount of information extractable from a noisy image are highly specific to the measurement method employed, there are general problems affecting each method in a similar way. In this section, we highlight the most relevant of these problems and seek to describe their level of systematic contamination of the ellipticity estimates. The tests are all carried out with the moment-based method {\scshape Deimos} \citep{Melchior11.1}, but great care has been taken to ensure that our approach and conclusions we draw from the tests can be generalized to other methods.

\begin{figure}
\includegraphics[width=\linewidth]{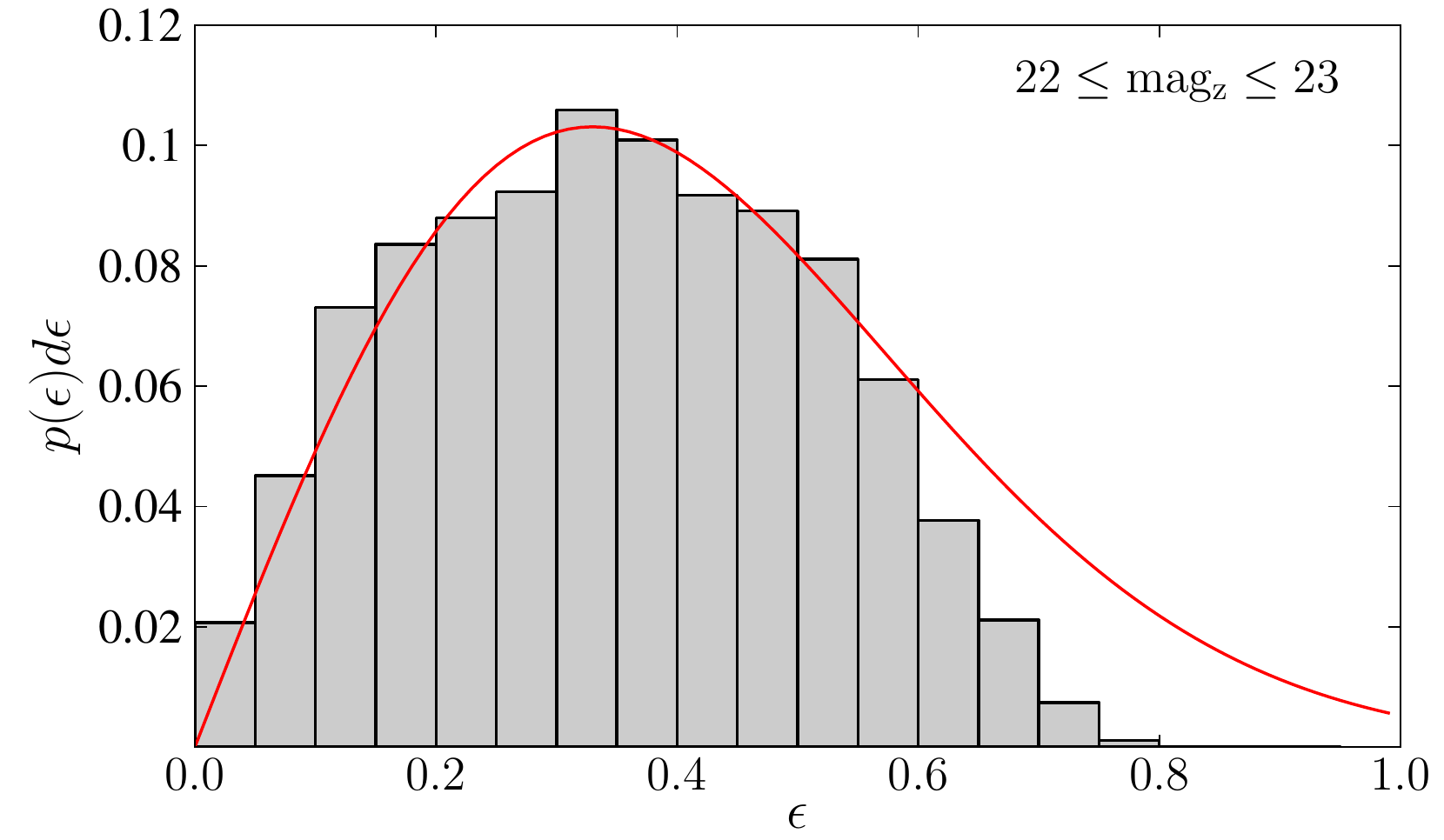}
\caption{Ellipticity distribution of GEMS galaxies from \citet{Haeussler07.1}. Galaxies in the specified magnitude range are only considered if {\sc Galfit} was able to fit a single-S\'{e}rsic model. For moderate ellipticities, the distribution is well described by the Rayleigh function \emph{(solid line)}, the expected distribution for the absolute value of the two-dimensional, or complex, ellipticity if each ellipticity component followed a common Gaussian distribution (here: $\sigma_e=0.33$).
The measured distribution lacks highly elliptical galaxies, either because galaxies with large ellipticity are not as abundant in nature as predicted by the Rayleigh distribution or because their measurement is more difficult (\autoref{sec:outliers} gives a possible explanation).}
\label{fig:gems_eps}
\end{figure}

The test galaxy images follow the S\'{e}rsic radial profile \citep{Sersic68.1} with intrinsic parameters, including the ellipticity, taken from model fits to galaxies in the GEMS {\it Hubble Space Telescope} survey \citep[see \autoref{fig:gems_eps}]{Haeussler07.1}. The galaxies are convolved with circular PSFs of Moffat-type \citep{Moffat69.1}. We choose two different PSF widths to mimic ground-based and space-based conditions. The pixel noise follows \autoref{eq:noise} with two levels of pixel noise, the first one corresponds to optimistic weak-lensing conditions of $\nu=35$, defined as in \citet[Equation 16 therein]{Erben01.1}, while the second one is one magnitude fainter ($\nu=15$). In both cases, the images exhibit prominent pixel noise, but the galaxies remain clearly detectable. The ground-based PSF has a Moffat-index of 9, the galaxy resolution factor $R_2\approx 0.4$ \citep[defined in][equation 8]{Hirata04.1}. The space-based PSF has a Moffat-index of 3, and the resolution is $R_2\approx 0.7$.

\subsection{Centroid errors}
\label{sec:centroiding}

Determining the accurate centroid $\bar{\mathbf{x}}$ of an object is crucial for any further shape analysis. This is obvious for moment-based approaches, for which the ellipticity definitions given in \autoref{eq:ellipticities} are sensible only if the centroid is chosen such that the dipole moment
\begin{equation}
\label{eq:dipole}
D_i \equiv \int d^2 x\ I(\mathbf{x}) \bigl(x_i - \bar{x}_i\bigr) = 0 \ \mathrm{for}\ i\in\lbrace1,2\rbrace,
\end{equation}
which therefore needs to be enforced by such methods.

Analogously, model-fitting approaches for galaxies or stars rely on models with peaked light distributions and finite support. Due to the limited amount of information in the image data, such models are often derived from a radial profile $p(r)$, whose radial coordinate undergoes an ellipticity transformation, rendering the light distribution axisymmetric. The radial coordinate is expressed relative to the centroid or the peak position of the light distribution,
\begin{equation}
r = \left |\, \begin{pmatrix}
1-\epsilon_1 & -\epsilon_2\\
-\epsilon_2 & 1+\epsilon_1
\end{pmatrix}
(\mathbf{x}-\bar{\mathbf{x}})\, \right |,
\end{equation}
such that the estimation of the model parameters $\epsilon_i$ explicitly depends on the estimation of $\bar{\mathbf{x}}$.

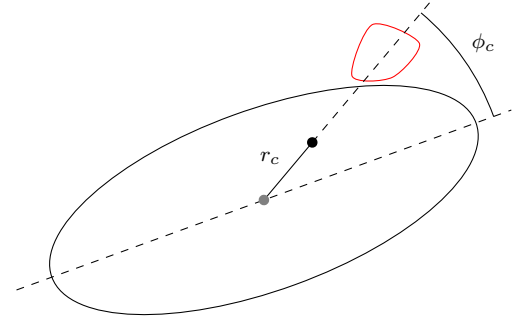
\begin{figure}
\begin{center}
\begin{tikzpicture}[scale=1,rotate=20]
\draw[dashed] (-3.5,0) -- (3.5,0); 
\draw (0,0) ellipse (3 and 1.2);
\draw[color=red] plot[smooth cycle] coordinates{(2.165,0.95)(2.655,1.25)(2.155,1.7)(1.665,1.15)};
\draw[rotate=30] (0,0) -- (1,0);
\draw[rotate=30, dashed] (1,0) -- (3.5, 0);
\fill[color=gray] (0,0) circle (0.7mm);
\fill[rotate=30,color=black] (1,0) circle (0.7mm);
\draw[thin] (3.25,0) arc (0:30:3.25cm);
\draw (3.139, 0.8411) node[anchor=south west] {$\phi_c$};
\draw (0.433, 0.25) node[anchor=south east] {$r_c$};
\end{tikzpicture}
\end{center}
\caption{Sketch of a perfectly elliptical "galaxy" affected by a positive noise fluctuation (red area), causing a centroid offset of length $r_c$ at an angle $\phi_c$ from the semi-major axis.}
\label{fig:centroid_sketch}
\end{figure}

With non-vanishing pixel noise, any measured centroid position is to some degree inaccurate, giving rise to an error $\Delta_{\bar{x}}$ in the measured location. It is useful to limit the discussion to perfectly elliptical shapes, i.e. galaxies whose isophotes have the same center, orientation, and ellipticity. By doing so, we can introduce polar coordinates, $\Delta_{\bar{x}} \rightarrow (r_c, \phi_c)$, and rotate again into a frame with only one non-vanishing ellipticity component  (cf. \autoref{fig:centroid_sketch}). Then, we can identify the centroid error with $\mathcal{P}$ in \autoref{eq:eps_distr},
\begin{equation}
\label{eq:centroiding_error}
p_c(\bepsilon^\prime | \bepsilon, \vec\theta, \nu) = \int_0^\infty dr_c \int_{-\pi/2}^{\pi/2} d\phi_c \ \bepsilon^\prime(r_c, \phi_c | \bepsilon, \vec\theta) \ p(r_c, \phi_c | \bepsilon, \vec\theta, \nu).
\end{equation}

\begin{figure}
\includegraphics[width=\linewidth]{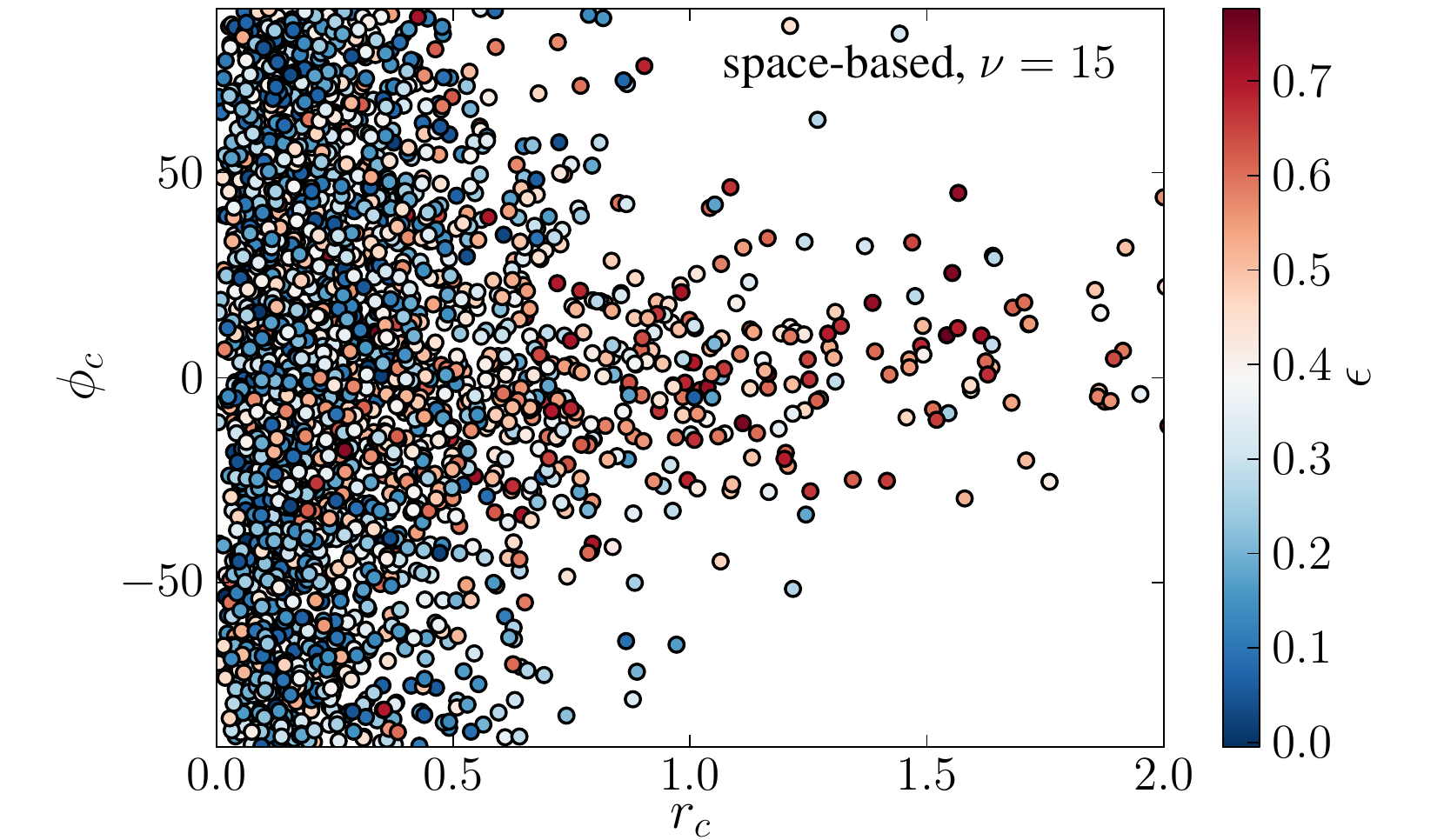}\newline
\includegraphics[width=\linewidth]{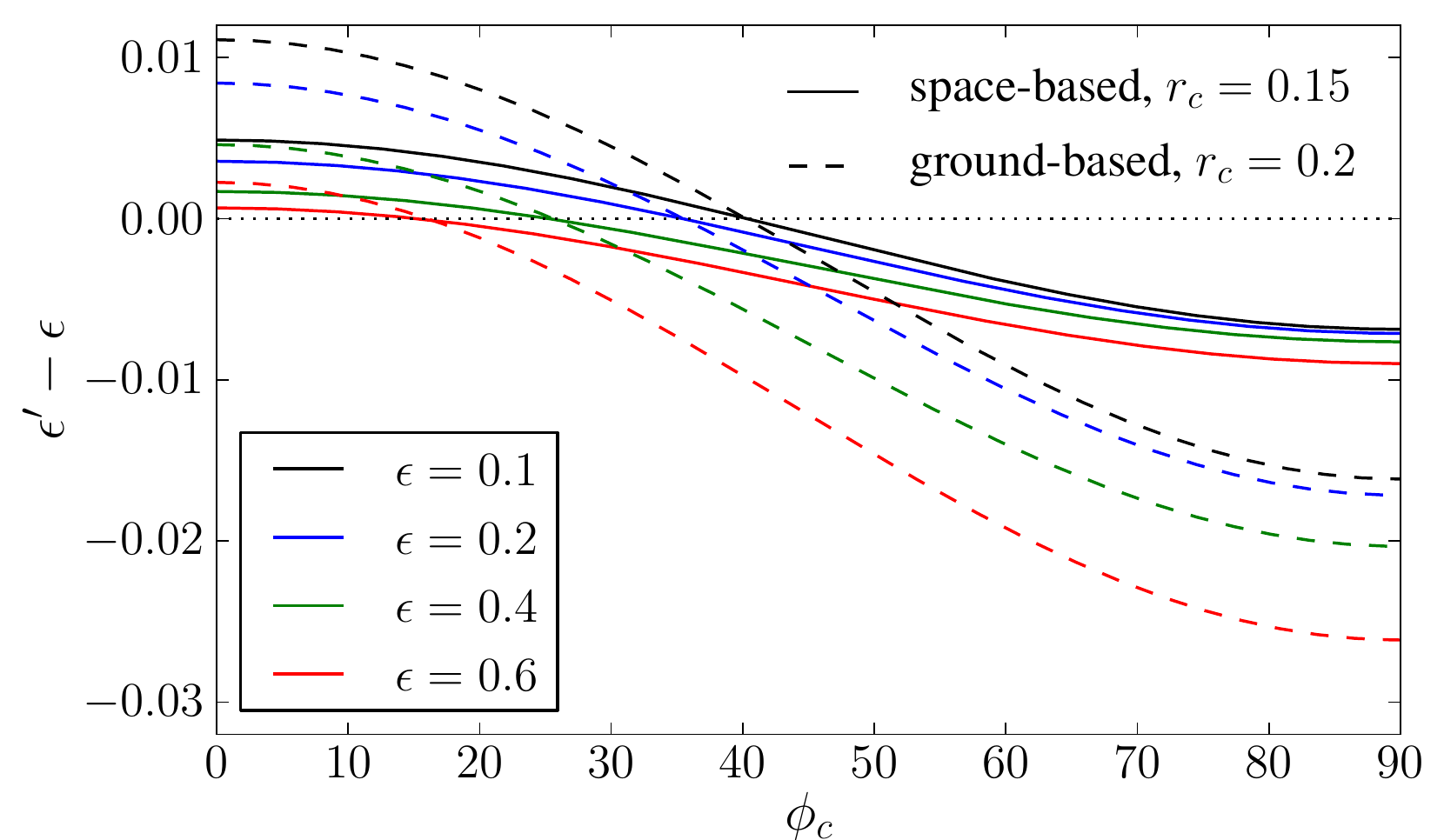}\newline
\includegraphics[width=\linewidth]{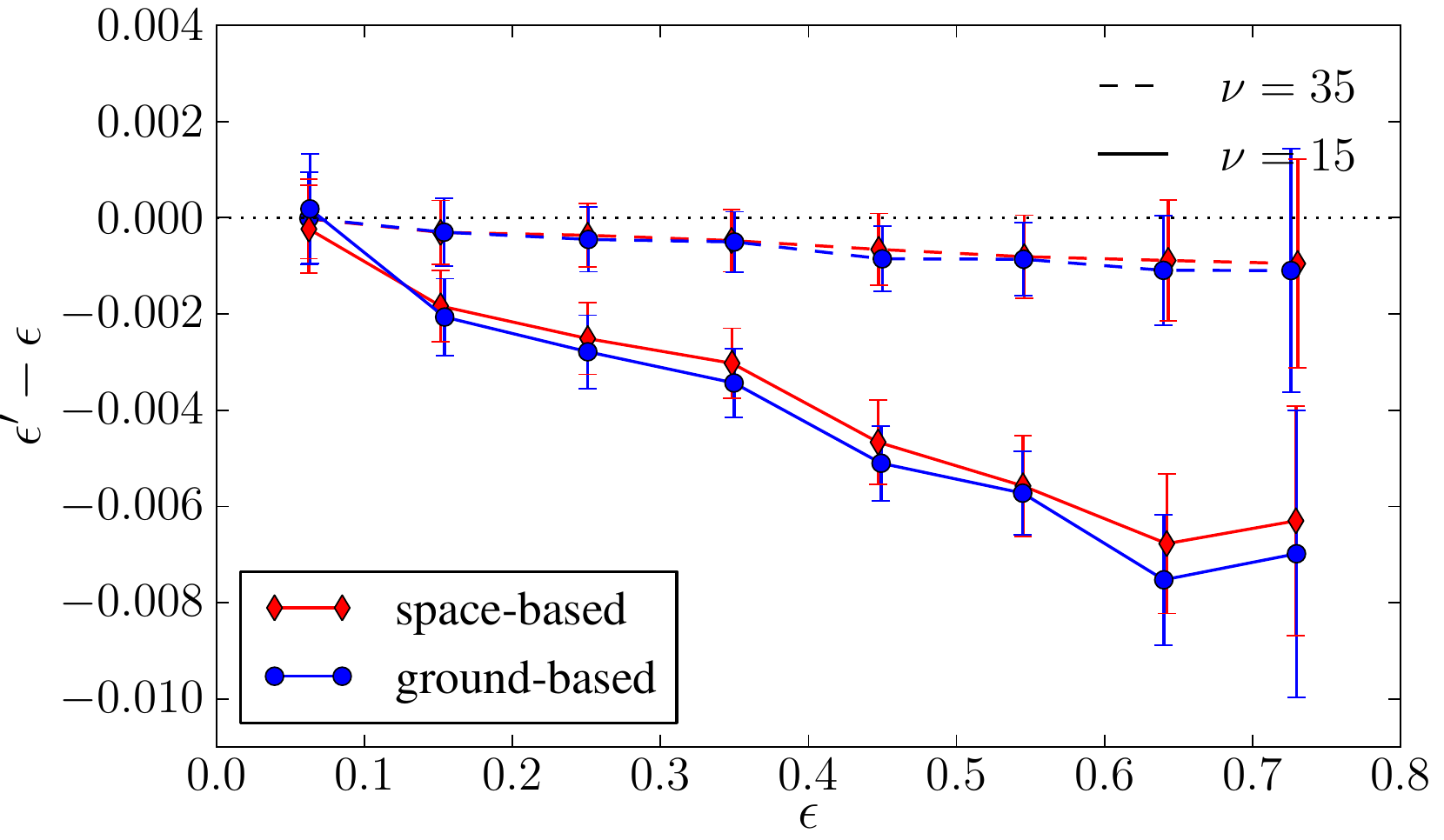}
\caption{\emph{Top:} Distribution of centroid offsets in polar $(r_c, \phi_c)$ coordinates for well-resolved galaxies with morphologies from the GEMS survey in images with high levels of pixel noise. \emph{Middle:} Ellipticity error as function of the miscentering angle $\phi_c$ for a noise-free galaxy image. The offsets $r_c$ were set to the median values of the high-noise ($\nu=15$) simulations. \emph{Bottom:} Ellipticity error as function of intrinsic ellipticity for the same ensemble used in the top panel, seen with two different noise levels (dashed and solid lines) and with space-based or ground-based resolution (red and blue lines). Error bars denote 1-$\sigma$ errors of the mean ellipticity in each bin. The scatter at the high ellipticity end is mainly driven by the lower number of simulated galaxies (cf. \autoref{fig:gems_eps})}
\label{fig:centroid_distribution}
\end{figure}

The top panel of \autoref{fig:centroid_distribution} shows the distribution of centroid errors under noise. In general, centroid errors for an elliptical light distribution show a similarly elliptical distribution, i.e. errors along the semi-major axis are larger than along the semi-minor axis  where the the light distribution has a larger gradient \citep[e.g.][]{Lewis09.1}. In the polar coordinate frame this translates into
\begin{equation}
\label{eq:centroid_distribution}
p(\phi_c, r_c) \propto r_c \exp\Bigl(-\frac{r_c^2}{a^2}\cos^2(\phi_c) - \frac{r_c^2}{b^2}\sin^2(\phi_c)\Bigr),
\end{equation}
where we assumed an elliptical Gaussian distribution for $p(\Delta_{\bar{x}})$ with semi-major axis $a$ and semi-minor axis $b$. Remarkable about this distribution is the enhanced probability of finding centroid errors with small $\phi_c$, and that this preferential alignment with the semi-major axis is most prominent for galaxies with large ellipticities. For this plot we chose the well-resolved galaxies with the high noise setting such as to emphasize the alignment of the centroid errors with the semi-major axis. For a ground-based instrument the distribution $p(\phi_c)$ would be much less peaked around zero for any $r_c$, and larger values of $r_c$ would be more likely. At lower noise, the distribution would be shifted towards smaller values of $r_c$.

To assess the impact of miscentering, we artificially shift the centroid position assumed in our shape measurement code, and record the resulting deconvolved ellipticity in absence of any pixel noise. The middle panel of \autoref{fig:centroid_distribution} illustrates the impact of miscentering on a typical disc-like galaxy as a function of the angle $\phi_c$. Shown are the ellipticity estimates for four different intrinsic ellipticities, with either space-based resolution (solid curves) or ground-based resolution (dashed curves). Several aspects of this plot are worth mentioning:
Centroid shifts with small angles $\phi_c$, i.e. aligned with the semi-major axis, shift the flux of the peak to non-vanishing distances without altering the orientation, thus leading to an increase of the ellipticity. On the opposite end, centroid shifts with large $\phi_c$ change the perceived orientation away from the actual one, thus lowering the inferred ellipticity. Furthermore, while over- and underestimate are fairly balanced for small $\epsilon$, galaxies with large ellipticity suffer much more strongly from miscentering.
Finally, the effects are stronger for the ground-based case since the PSF deconvolution amplifies any ellipticity signal, both the true and the spurious one. This is additionally enhanced by the larger offsets $r_c$ encountered for the ground-based resolution at fixed $\nu$.

The shape of the miscentering curves can be approximately described by 
\begin{equation}
\label{eq:error_centroiding}
|\bepsilon^\prime| \approx \bigl[|\bepsilon| + \epsilon_0 -\epsilon_{90}\bigr] \cos^2(\phi_c) + \epsilon_{90},
\end{equation}
where the parameters $\epsilon_{0,90}\geq0$ depend on $r_c$ (which in turn depends on $\nu$) and $\epsilon$, as well as other source parameters $\vec\theta$, most notably the width of the PSF, and the method employed.

In the bottom panel of \autoref{fig:centroid_distribution} we show the ellipticity error induced by miscentering. We can see that the overall effect of the miscentering is a small, but consistent underestimation of the inferred ellipticity, which scales linearly with $\epsilon$. This is not surprising, since large ellipticities suffer more strongly from miscentering. This effect is partially compensated by the preferential alignment of the centroid with the semi-major axis such that the centroid errors are not entirely isotropic and the average error bias is smaller than what would naively be expected when only considering the middle panel of \autoref{fig:centroid_distribution}.
Qualitatively, this result is in good agreement with the derivation by \citet[see their section 8.2 and Equation 8.1]{Bernstein02.1} in that the bias depends linearly on $\epsilon$ and scales approximately as $\nu^{-2}$. However, we only find a mild dependence on the image resolution $R$.

\subsection{Misalignment}
\label{sec:misalignment}

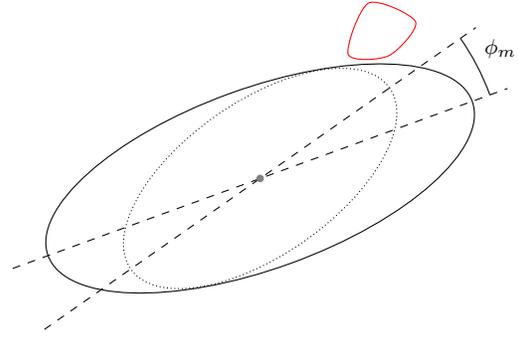
\begin{figure}
\begin{center}
\begin{tikzpicture}[scale=1,rotate=20]
\draw[dashed] (-3.5,0) -- (3.5,0); 
\draw (0,0) ellipse (3 and 1.2);
\draw[color=red] plot[smooth cycle] coordinates{(2.165,0.95)(2.655,1.25)(2.155,1.7)(1.665,1.15)};
\draw[rotate=15, dashed] (-3.5,0) -- (3.5,0);
\draw[densely dotted, rotate=15] (0,0) ellipse (2.1 and 1.02);
\fill[color=gray] (0,0) circle (0.5mm);
\draw[thin] (3.25,0) arc (0:15:3.25cm);
\draw (3.222, 0.424) node[anchor=south west] {$\phi_m$};
\end{tikzpicture}
\end{center}
\caption{Sketch of a perfectly elliptical "galaxy" (solid) affected by a positive noise fluctuation (red area), which gives rise to a misalignment of an elliptical model or weight function with respect to the true orientation by an angle $\phi_m$ (dotted).}
\label{fig:misalignment_sketch}
\end{figure}

In noisy images, methods that employ either an elliptical model or -- in the case of moment-based approaches -- an elliptical weight function are subject to random errors in the determination of the orientation. Like before, we choose a coordinate system aligned with the semi-major axis of the elliptical source and introduce the misalignment angle $\phi_m$ (cf. \autoref{fig:misalignment_sketch}). Identifying misalignment as the conceptual problem in \autoref{eq:eps_distr} leads to
\begin{equation}
\label{eq:ellipticity_error}
p_m(\bepsilon^\prime | \bepsilon, \vec\theta, \nu) = \int d\phi_m \ \bepsilon^\prime(\phi_m | \bepsilon, \vec\theta) \ p(\phi_m | \bepsilon, \vec\theta, \nu),
\end{equation}
where $\bepsilon^\prime(\phi_m | \bepsilon, \vec\theta)$ denotes the ellipticity measurement performed with a potentially erroneous orientation. As indicated in \autoref{fig:misalignment_sketch}, we expect the measured shape of the object to be biased towards smaller sizes and ellipticities. Estimates with smaller sizes were indeed reported by \citet[Figure 1 therein]{Kacprzak12.1} for model-fitting approaches.

\begin{figure}
\includegraphics[width=\linewidth]{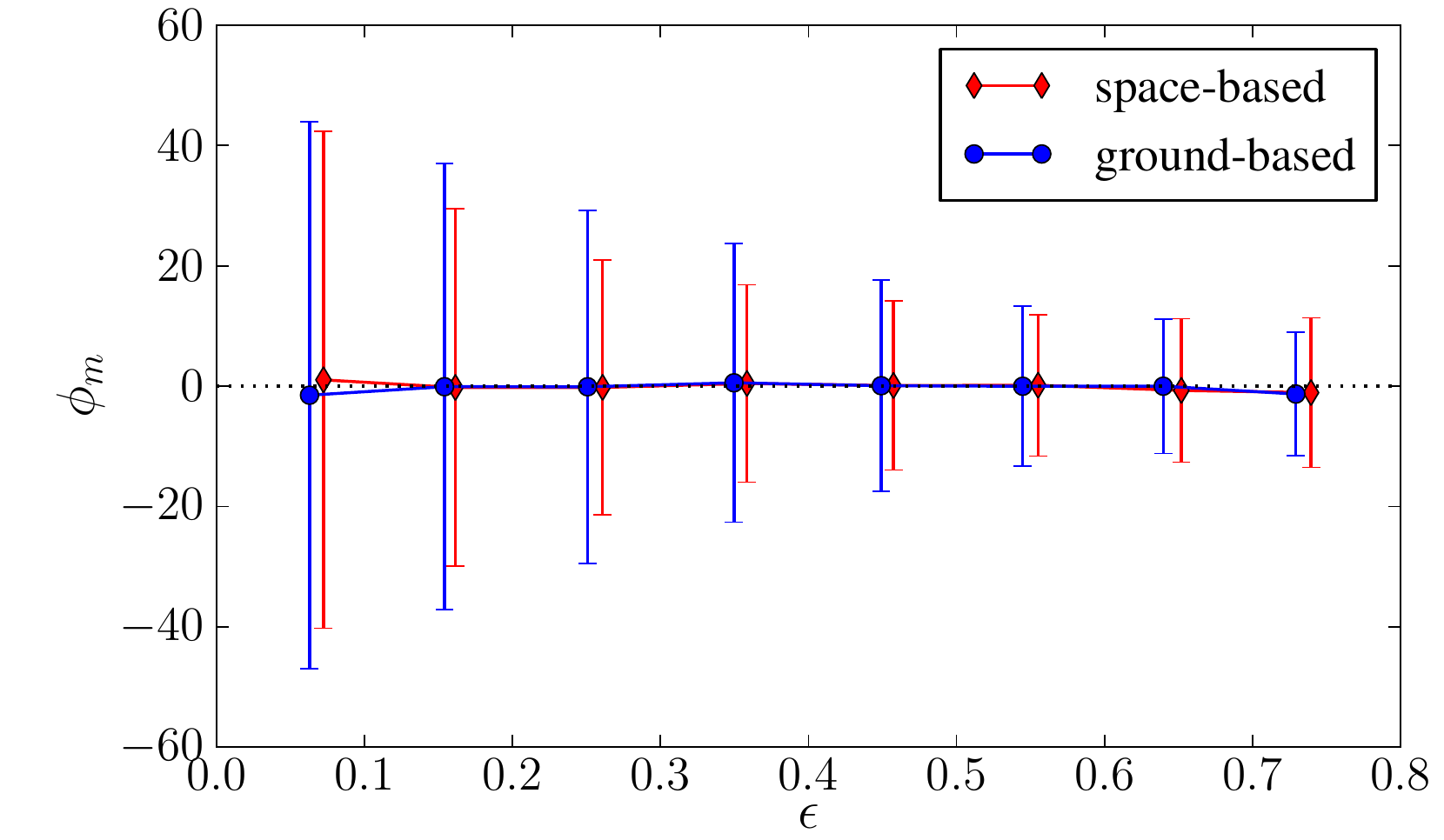}\\
\includegraphics[width=\linewidth]{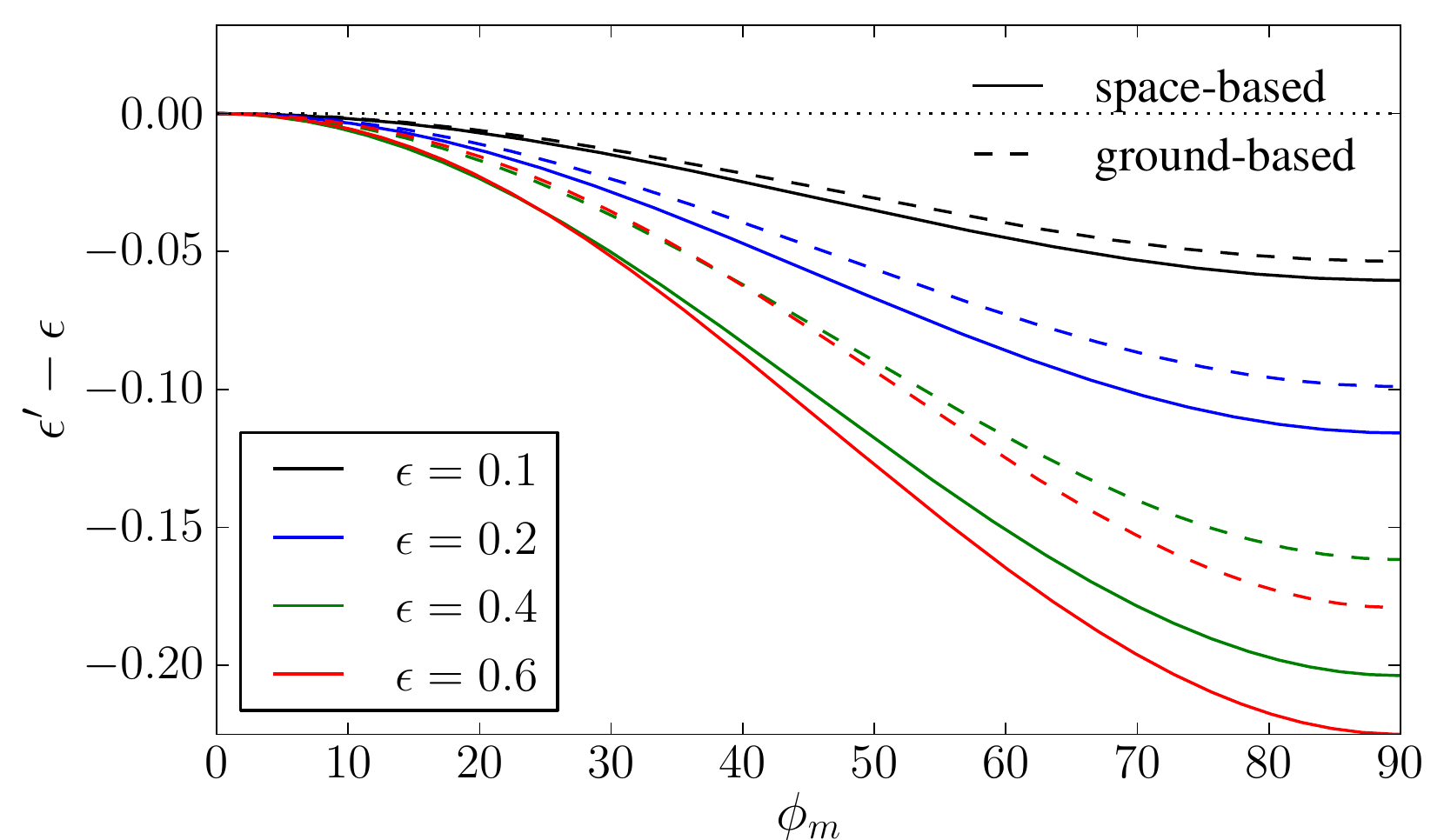}\\
\includegraphics[width=\linewidth]{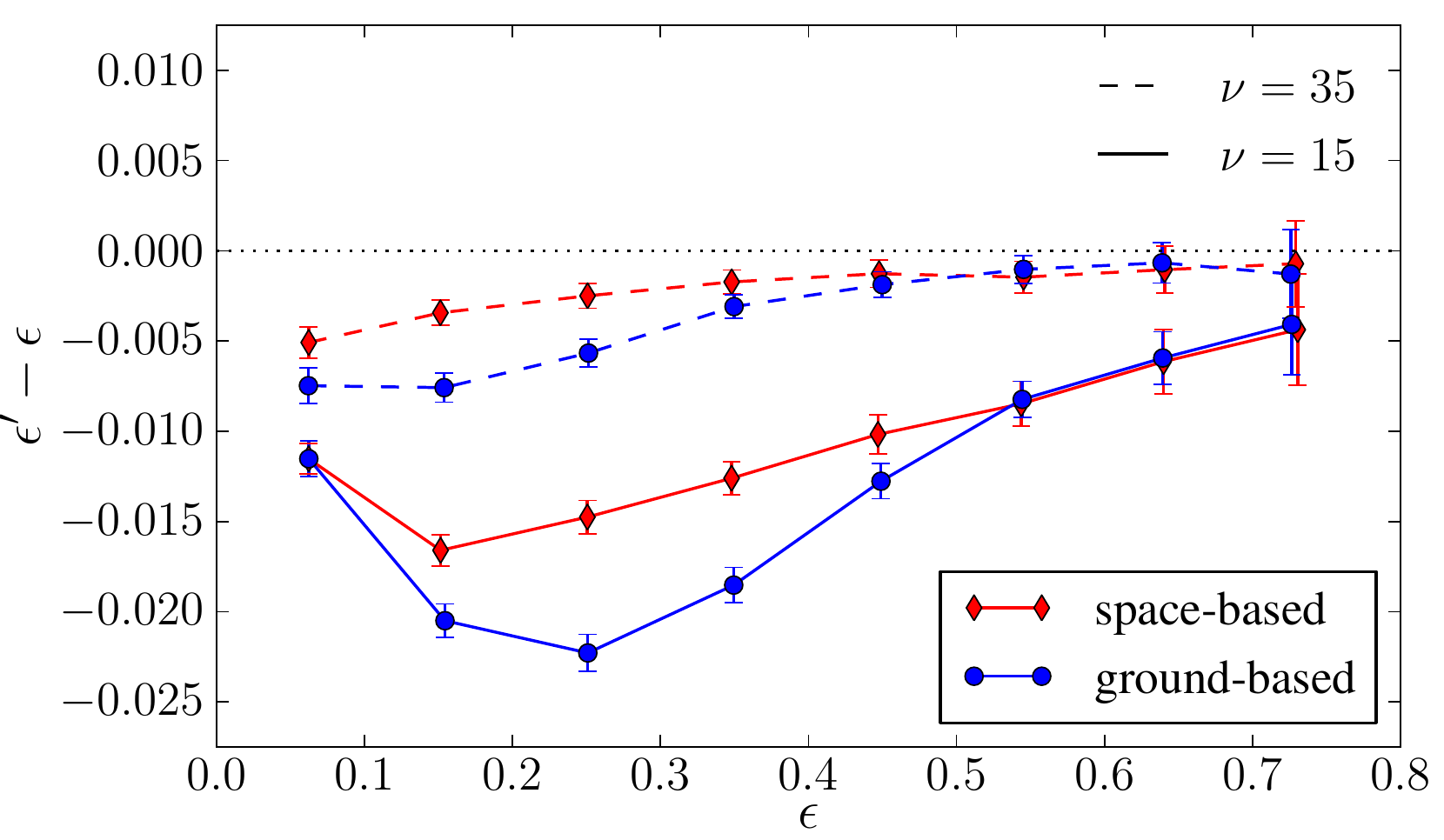}
\caption{\emph{Top:} Width of the orientation error angle $\phi_m$ for the high noise level. For better visibility, the space-based data has been shifted horizontally by 0.01. Error bars denote 1-$\sigma$ intervals. \emph{Middle:} Error of the ellipticity when the weight function is rotated by the angle $\phi_m$ for four different galactic ellipticities. 
\emph{Bottom:} Error of the ellipticity in noisy images caused by misalignment as a function of galactic ellipticity. Error bars denote the 1-$\sigma$ errors of the mean in each ellipticity bin.}
\label{fig:misalignment_distr}
\end{figure}

In \autoref{fig:misalignment_distr} we show -- from top to bottom --  the measured distribution of misalignment angles $\phi_m$ from noisy images, the impact of misalignment on the ellipticity estimates in noise-free images, and the net effect of misalignment on the inferred ellipticity as a function of the true ellipticity. For these tests, $\phi_m$ denotes the orientation of the weight function used for the measurement, but since the PSF is circular, the orientation is identical to the actual orientation of the galaxy. That means, $\bepsilon^\prime(\phi_m)$ is obtained by rotating the correctly matched elliptical weight function by the angle $\phi_m$.\footnote{In order to single out the effect of misalignment, we do not alter the size nor the total ellipticity of the weight function, even though these changes would occur in practice.}
Two essential finding can be made with this test. Pixel noise leads to an error on the orientation inferred from the image data, which becomes smaller with increasing ellipticity. This is not surprising since a circular source has a uniform distribution of orientation angles, while increasing the ellipticity renders the determination of the orientation easier, even in noisy images. The second finding is that for any $|\phi_m| >0$ an underestimation of $\bepsilon$ is observed with a characteristic shape that approximately follows the relation
\begin{equation}
|\bepsilon^\prime| \approx (|\bepsilon| - \epsilon_{90}) \cos^2(\phi_m) + \epsilon_{90},
\end{equation} 
with some $\epsilon_{90}$ that depends on the parameters $\vec\theta$ of the apparent galactic shape. Combining both findings, we obtain a low bias induced in noisy images caused by misalignment. Looking at the bottom panel of \autoref{fig:misalignment_distr}, we can see that the effects of misalignment are more severe for modestly elliptical galaxies, where we observe an underestimation also in the moderate noise case with $\nu=35$. This is an immediate consequence of the poor alignment constraints for galaxies with small $|\bepsilon|$, and as misalignment is dependent on the apparent, i.e. convolved, ellipticity, ground-based imaging is more prone to this kind of bias.

\subsection{Method-related limitations}
\label{sec:models}

So far, we have dealt with the problem of pixel noise assuming that a method for estimating the ellipticity is employed, whose validity does not become questionable in presence of noise. Unfortunately, we cannot expect shape measurement methods to be entirely indifferent to increasing amounts of noise.

Moment-based methods, such as KSB \citep{Kaiser95.1}, {\sc Holics} \citep{Okura09.1} or {\sc Fdnt} \citep{Bernstein10.1}, need to apply a weight function to the data to suppress the impact of pixel noise at large distance from the source. The application of the weight function leaves an imprint on the measured moments, which can be corrected, albeit only approximately. With increasing amplitude of the noise, the only way to limit the variance of the ellipticity estimates is to shrink the weight function. Then, the approximation for the weight function correction become increasingly inaccurate, leading to errors in the deweighted moments. The direction and amplitude of the errors are a function of the apparent galaxy shape, in particular its slope, and properties of the PSF \cite[e.g.][Figure 1 therein]{Melchior11.1}.

Model-fitting approaches do not need an artificial weight function as they make use of the compactness of their galaxy model, which is often related to the S\'{e}rsic radial profile. One problematic aspect is the validity of this model -- or model family -- to faithfully describe the morphologies of all galaxies present in the observation. If data with a higher significance and possible also a higher spatial resolution is available, the model assumptions can be verified, but this is often unfeasible. Bayesian approaches in model-fitting \citep[e.g. {\scshape Lensfit},][]{Miller07.1, Kitching08.1} additionally employ priors on some parameters of the model, which are hard to estimate from the data alone. These priors are themselves derived from data of higher quality, and do not necessarily apply to the data at hand.
But even if galaxies were purely elliptical -- as many models assume -- unbiased shear estimates require the radial profile to be accurately matched to the observed galaxies \citep{Voigt10.1}, which cannot be guaranteed with images of severely limited significance.

Model assumptions could be avoided by using a decomposition into complete basis function sets, such as shapelets \citep{Refregier03.1} or S\'{e}rsiclets \citep{Ngan09.1}. 
However, the pixel noise limits the number of modes used in the fit such that the resulting model becomes dominated by the shape of the zeroth order. In case of the circular shapelet basis function set, this introduces a bias towards circular objects \citep{Melchior10.1}. For S\'{e}rsiclets, which can be considered a generalization of shapelets, using a finite number of modes leads to a relation between the slope of the radial profile and the spatial scale of higher-order fluctuations, which is not necessarily obeyed by observed galaxies \citep{Andrae11.2}.

In summary, any method, which deals with severely degraded data invokes additional assumptions about the data, which may turn out to be wrong once an adequate assessment can be devised, e.g. with data of higher quality. It needs to be shown that the methods employed are able to meet requirements demanded to reach the scientific goals of the project at hand.
Therefore, simulations with simplified galaxy models \cite[e.g.][]{Heymans06.1,Bridle10.1,Kitching12.1} provide a clean way of comparing several methods,
but should be complemented by simulations with realistic galaxy morphologies \citep[such as][]{Massey04.1, Meneghetti08.1, Mandelbaum12.1}. 

Additional problems arise from the occurrence of outliers, i.e. $|\bepsilon^\prime|\geq1$. From the discussion in \autoref{sec:error_propagation} and by looking at \autoref{fig:MarsagliaBias}, we know that outliers must occur if the combination of apparent ellipticity and pixel noise exceeds some value. Moment-based methods will just let these outliers pass (unless they try to avoid them by shrinking the weight function) and expect a subsequent lensing analysis to deal with them. We revisit this problem in \autoref{sec:outliers}. On the other hand, by construction model-fitting methods cannot have such outliers, but they will encounter catastrophic events, such as convergence failures, and flag these objects, which again excludes them from the analysis later on.

Alternatively, model-fitting approaches might invoke a prior on the ellipticity, which could in principle completely prevent the occurrence of outliers or modeling failures. However, this comes at the price of essentially recovering the prior for data with very little constraining power on the ellipticity. As we said above, this prior does not necessarily describe the actual data accurately. Moreover, even if the prior is an accurate description of the ensemble ellipticity distribution, the fact that at fixed noise level outliers preferentially occur for galaxies with large ellipticity leads to a stronger impact of the prior on those galaxies. Since large ellipticities are less abundant than those around $|\bepsilon|=0.3$ (cf. \autoref{fig:gems_eps}), we expect a bias towards the prior which grows with galaxy ellipticity. 

For years, attempts have been made to calibrate the bias away based on simulated images, e.g. by introducing global "fudge" factors that boost the measured ellipticities \citep[e.g.][Table A1 therein]{Heymans06.1}. More recently, sophisticated supervised learning methods have been employed to correct for the bias as a function of several input parameters \citep{Gruen10.1,Tewes12.1,Kacprzak12.1}. Although they acknowledge and account for the presence of biases in the measurements, irrespective of their origin, they hinge on quality, size, and representative nature of the training set. We thus regard these methods to work well for galaxies that are abundant in the training data \citep[cf.][]{Kitching12.1} and whose properties accurately resemble those of actually observed galaxies. 

\section{Inadequate shear statistics}
\label{sec:statistics}

We now turn to a different type of systematic problem caused by the pixel noise, which does not occur at the level of individual ellipticity estimates but later on, namely when statistics of the measured ellipticity distribution are calculated. Pixel noise can render these statistics difficult to interpret or entirely inappropriate by violating their fundamental assumptions. One such assumption in virtually every lensing analysis is that a proper definition of ellipticity provides an unbiased estimator of the (reduced) shear,
\begin{equation}
\label{eq:unbiased_estimator}
\langle\bepsilon\rangle \ \equiv \int d^2\epsilon\ \bepsilon\, p(\bepsilon) = \mathbf{g}.
\end{equation} 
For a noise-free measurement of the ellipticity in the form of \autoref{eq:ellipticity}, this is in fact the case \citep{Seitz97.1,Bartelmann01.1}.

For what follows, we are going to assume the optimistic scenario, in which the measurement methods provide ellipticity estimates, which follow directly the Marsaglia distribution of \autoref{eq:marsaglia}. That means all complications discussed in \autoref{sec:measurements} are eliminated. In practice, this can be realized by using the sampling method outlined in \autoref{sec:sampling}, which provides a realistic ellipticity distribution of Gaussian-shaped galaxies under noise but is not affected by shape-measurement biases. We recall the two main distinctions between the two popular moment-based ellipticity estimators from \autoref{sec:eps_chi}: $\bepsilon$ has a problematic distribution under noise (cf. \autoref{fig:eps_planes}), while $\bchi$ depends on the shear in a non-linear way. Both features will prove to be problematic.

\subsection{Outliers and their rejection}
\label{sec:outliers}
\begin{figure}
\includegraphics[width=\linewidth]{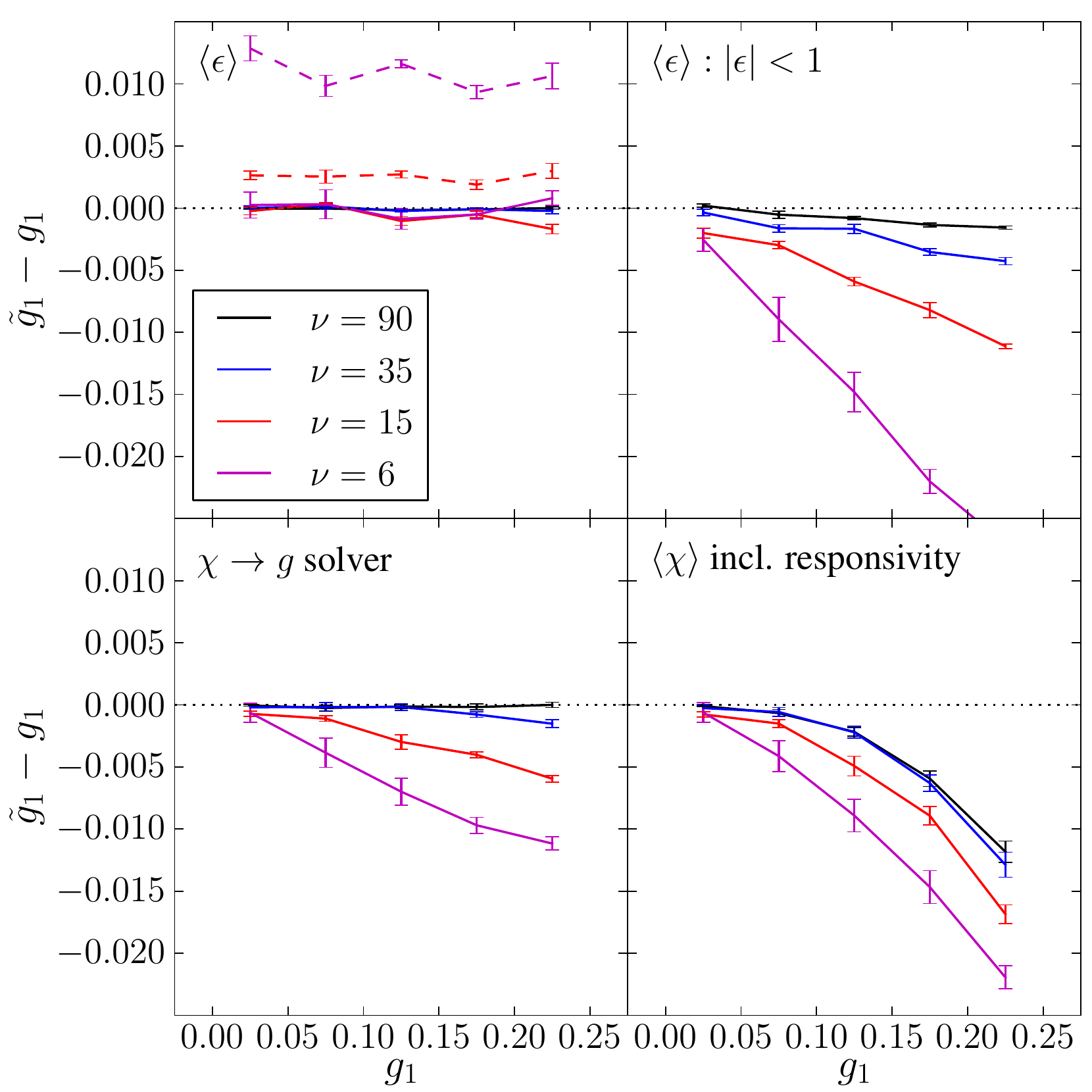}
\caption{Performance of shear statistics under pixel noise. \emph{Top left:} Average of the entire $\bepsilon$ distribution, following \autoref{eq:unbiased_estimator}. \emph{Top right:} Same as in the left panel, but after rejection of outliers with $|\bepsilon|\geq1$. \emph{Bottom left:} Non-linear solver for the shear from measurements of $\bchi$, based on \autoref{eq:chi_g}. \emph{Bottom right:} Linearized relation between $\bchi$ and the shear, following \autoref{eq:mean_e}. For all panels, shear was only applied on 1-direction (solid lines; dashed lines in the top left panel: $g_2=0.1$), and the ellipticity noise was of Marsaglia-type, simulated with the algorithm described in \autoref{sec:sampling}. Means and errors are taken from 10 independent noise realization at each value of the shear, each realization comprised 10,000 samples.}
\label{fig:stats}
\end{figure}

As we have stressed several times now, the pixel noise can lead to ellipticity outliers with $|\bepsilon^\prime| \geq 1$ (cf. \autoref{fig:MarsagliaBias}) and thus gives rise to a second population of ellipticity measurements, namely those on the unit circle in $\bepsilon$-space. If the resulting sample is naively inserted in \autoref{eq:unbiased_estimator}, the presence of these outliers, whose position on the ring is still loosely correlated with their noise-free location, should lead to an overestimation of the inferred shear, simply because they all have unit ellipticity and thus large impact on the mean of the distribution. Weirdly, this is \emph{not} observed in the top-left panel of \autoref{fig:stats}, where we show the mean ellipticity as a function of shear and noise level. If the shear only has one non-vanishing component, the population on the unit circle almost perfectly compensates the underestimation we expect from the mean of the Marsaglia distribution (cf. \autoref{fig:MarsagliaBias}). This balance is delicate, and it is quite possible that it is not maintained for galaxies with a different radial profile, where the means and errors of the measured moments deviate from our Gaussian calculation. But even if the balance persisted, statistics other than the mean, e.g. the two-point correlation function, will in general pick up the presence of outliers and perform erratically. Another concerning aspect of $\langle\bepsilon\rangle$ as shear estimator becomes apparent when the other shear component comes into play. Because the noise affects the outliers by altering their phase on the unit circle, the estimate $\tilde{g}_1$ becomes dependent on $g_2$  (dashed lines). The cross-talk between the two shear components is relevant for the faintest two noise settings, where the outliers constitute a significant portion of the entire sample.\footnote{The other estimators shown in \autoref{fig:stats} do not suffer significantly from this sort of cross-talk since they either do not have outliers (top right panel) or the outlier population has a well-behaved shape (both estimators based on $\bchi$). Hence, we only show the effect for $\langle\bepsilon\rangle$.}

On the other hand, excluding these outliers would be absolutely justified since any ellipticity definition needs to be bound by 1, otherwise the ratio of semi-minor to semi-major axis is non-sensical. Unfortunately, this commonly adopted approach also leads to biases because we now sample from
\begin{equation}
p_o(\bepsilon^\prime|\bepsilon,\vec\theta,\nu) = \int_{|\bepsilon+\mathbf{n}|<1} d^2n\ (\bepsilon + \mathbf{n})\ p(\mathbf{n}|\bepsilon,\vec\theta,\nu).
\end{equation}
Even if $p(\mathbf{n})$ were isotropic and had zero mean, the truncation of the integration range at the ellipticity unit circle poses a bias for all galaxies with sufficiently large ellipticity. If $p(\mathbf{n})$ is monotonically decreasing with increasing $|\mathbf{n}|$ as it is common, this bias will be negative because the outlier-excluded distribution is more compact than the actual noisy distribution. In other words, this negative bias increases with ellipticity or, equivalently, with the probability of obtaining a noise contribution that can push the measured ellipticity beyond the unit circle.  Consequently, sampling from $p_o(\bepsilon^\prime|\bepsilon)$ rather than from $p(\bepsilon)$ in \autoref{eq:unbiased_estimator} results in a low bias on $\mathbf{g}$ since any coherent distortion of the ellipticity distribution will increase the probability of falling outside of the unit circle for all galaxies, whose intrinsic ellipticity is aligned with the additional distortion. Hence, this distortion becomes suppressed in the outlier-rejected sample. This is shown in the top-right panel of \autoref{fig:stats}. The bias becomes more prominent with increasing shear since large ellipticities become more abundant and the entire distribution thus more likely to create outliers at any non-vanishing noise level.

It is important to note that even though we have discussed the outlier problem solely in terms of moment-based measurements, shear estimates from model-fitting methods are susceptible to the same bias: Any filtering on catastrophic modeling failures is equivalent to outlier rejection, and if such failures become more prominent with increasing source ellipticity, the shape of the bias curves will follow the one in the top-right panel of \autoref{fig:stats}.

\subsection{Non-linear statistics}
\label{sec:non-linear-statistics}

Using $\bchi$ rather than $\bepsilon$ provides the advantage of a much simpler noise distribution with a closed description: the Marsaglia distribution of \autoref{eq:marsaglia}. However, the relation between $\bchi$ and $\mathbf{g}$ is non-linear. As before, we are thus faced with error propagation in a non-linear system and have to expect biased shear estimates even from a perfect measurement. 

The theoretically correct way of obtaining shear estimates from measurements of $\bchi$ is solving for the shear $\mathbf{g}$ that nulls the mean of the source-plane ellipticity \citep{Schneider95.1},
\begin{equation}
\bchi^s = \frac{\bchi - 2\mathbf{g}+\mathbf{g}^2\bchi^*}{1+|\mathbf{g}|^2 - 2\mathbb{R}(\mathbf{g}\bchi^*)}
\label{eq:chi_g}
\end{equation}
as the distribution of unlensed galaxy ellipticities is assumed to be isotropic, i.e. with zero mean. To our knowledge, the bottom-left panel of \autoref{fig:stats} shows the first application of this estimator. While unbiased for all shears at zero noise, and being still unbiased for $\nu=35$ and $g_1 < 0.15$, the solver exhibits negative bias for the two faintest noise settings.

In practice, \autoref{eq:chi_g} is approximated to first order in the shear, which leads to \autoref{eq:mean_e} and the usage of the so-called responsivity correction. Unsurprisingly, this simplified relation introduces its own bias that grows with the shear, which is shown in the bottom-right panel of \autoref{fig:stats}. Moreover, this simplified estimator does not perform any better than the fully non-linear solver in the left panel. In fact, our novel estimator appears to cope well with outliers in the noisy ellipticity distribution and to perform more reliably than the other three. It has the highest computational demands as it involves the minimization of an objective function, the modulus of $\bchi^s$ in \autoref{eq:chi_g}, but can be implemented efficiently.

\subsection{Ellipticity weights}
\label{sec:weights}

Often, ellipticity estimates do not directly enter \autoref{eq:unbiased_estimator}, but get weighted before. Such a weighted average is tempting for two reasons. First, one can reduce the noise-induced variance by penalizing faint galaxies. Second, one might even be able to reduce systematic biases by down-weighting galaxies from regimes, where the employed method yields consistently wrong ellipticity estimates. For instance, as the outlier problem increases with apparent ellipticity (cf. \autoref{sec:outliers}) while the responsivity to shear becomes weaker, applying larger weights for less elliptical galaxies seems logical.

\begin{figure}
\includegraphics[width=\linewidth]{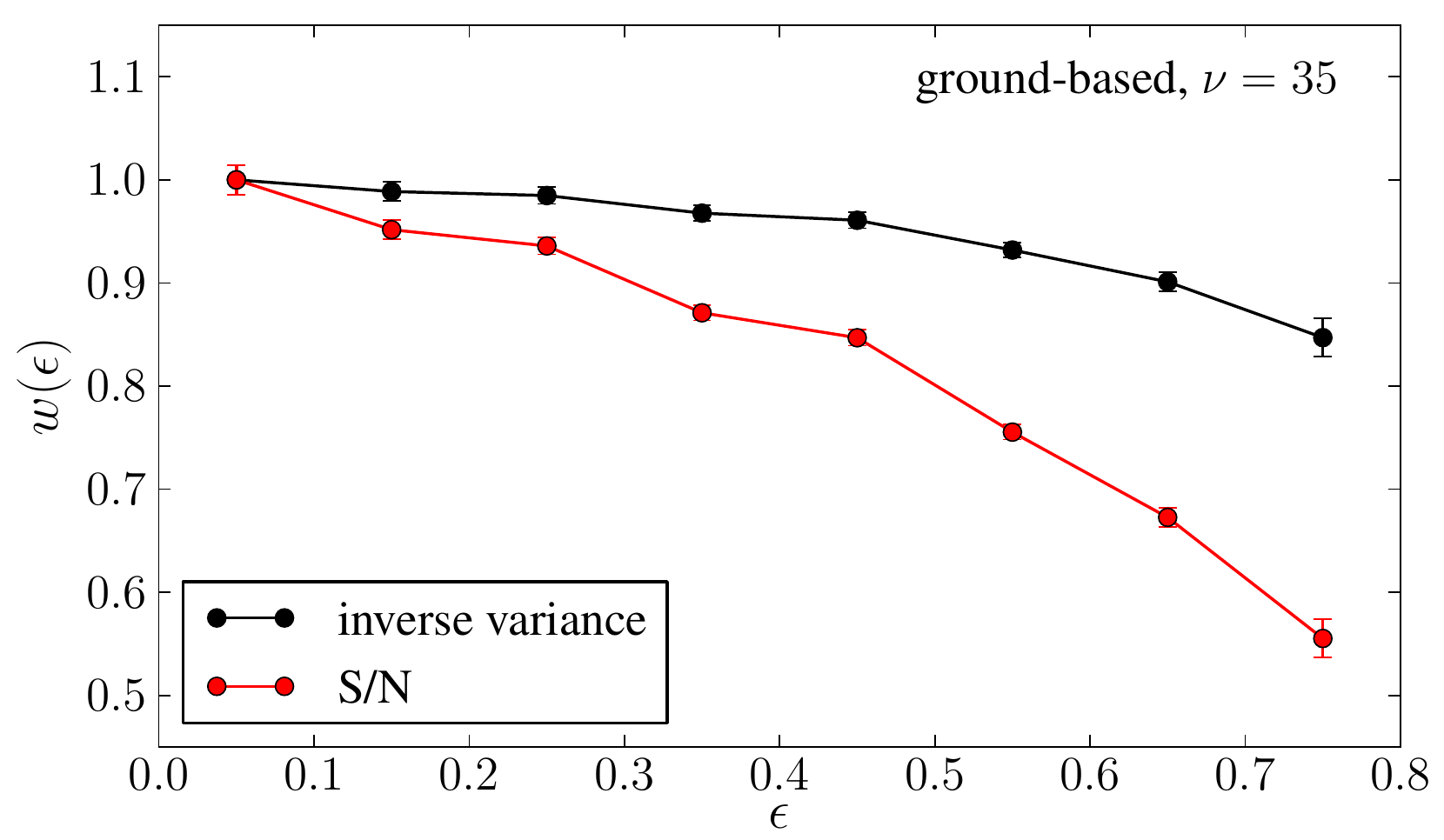}
\caption{Ellipticity weights: \emph{Inverse variance} denotes the scheme that attempts to minimize the mean square error, considering the ellipticity-dependent measurement noise $\sigma_n$ and the intrinsic scatter $\sigma_e$, according to $w\propto[\sigma_n^2 + \sigma_e^2]^{-1}$ \citep{Hoekstra00.1}. \emph{S/N}  denotes a scheme where the weight is directly proportional to the measurement significance, which often favors circular objects. Measurement errors and $S/N$ are obtained from the {\sc Deimos} method, the intrinsic dispersion was assumed to be $\sigma_e=0.3$.}
\label{fig:weights}
\end{figure}

But a weighting scheme can introduce a bias on its own, even for perfectly unbiased, noise-free ellipticity estimates, namely in this seemingly beneficial case of ellipticity-dependent weights. If we replace the theoretically unbiased average $\langle\bepsilon\rangle$ from \autoref{eq:unbiased_estimator}
by its weighted and properly normalized equivalent
\begin{equation}
\label{eq:weighted_av}
\langle w\bepsilon \rangle =\frac{\int d^2\epsilon\ w\bepsilon\, p(\bepsilon)}{\int d^2\epsilon\ w\, p(\bepsilon)},
\end{equation}
we can immediately see that the result is the same, i.e. unbiased, only if $w$ does not depend on $\bepsilon$. If the weight decreases with increasing ellipticity, we are confronted with an altered and more compact distribution
\begin{equation}
p_w(\bepsilon^\prime) = w(\bepsilon^\prime) p(\bepsilon^\prime),
\end{equation}
similar to the case of the outlier-rejected distribution.
Even if we do not explicitly want the weighting scheme to penalize large ellipticities, the measurement method might report statistics of the effective signal-to-noise level $\nu$ that are turned into weights. Due to the more difficult task to measure large ellipticities, these statistics tend to implicitly depend on the source ellipticity (cf. \autoref{fig:weights} for two plausible weighting schemes). For the simple case of an approximately linear relation of the weight on $\bepsilon$, described by the intercept $w_0$ and the slope $c$, we show in \autoref{sec:weighting_bias} that the resulting bias on the shear estimate is to first order purely multiplicative,
\begin{equation}
\label{eq:weighting_bias}
\langle w\bepsilon \rangle - \langle \bepsilon \rangle \approx \mathbf{g} \sqrt{\frac{2}{\pi}}\frac{c}{w_0}\sigma_e,
\end{equation}
where $\sigma_e$ denotes the dispersion of the ellipticity estimates. With reasonable values of $\tfrac{c}{w_0}\approx -0.1$, we obtain a bias of $-2.5\%$.

\begin{table*}
\begin{minipage}{\linewidth}
\caption{Overview of the ellipticity and shear estimation biases described in this work.}
\label{tab:summary}
\begin{tabular}{llll}
\hline
Description & Section & Effect & References\\
\hline
Error propagation & \autoref{sec:error_propagation} & noise distribution shifted and skewed, average biased low  & this work\\
& & strongly ellipticity-dependent & \\
Centroiding error & \autoref{sec:centroiding} & bias $\propto \epsilon/\nu^2$ & \citet{Bernstein02.1}\\
Misalignment & \autoref{sec:misalignment} & bias most problematic for intermediate ellipticities & this work\\
Various shape estimation issues & \autoref{sec:models} & deviation of measured ellipticity distribution from its expectation & see \autoref{sec:models}\\
Ellipticity outliers & \autoref{sec:outliers} & Cross-talk between shear components (with outliers)  & this work\\
& & or multiplicative shear understimation (after outlier rejection) & \\
Non-linear shear statistics & \autoref{sec:non-linear-statistics} & Multiplicative shear underestimation & this work\\
Ellipticity-dependent weighting & \autoref{sec:weights} & Multiplication shear underestimation, negligible in correlation function $\xi_g$ & this work\\
\hline
\end{tabular}
\end{minipage}
\end{table*}

Also the two-point correlation function of (post-lensing) ellipticities
\begin{equation}
\label{eq:epseps}
\langle \bepsilon_i \bepsilon_j\rangle (\theta) = \xi_g(\theta),
\end{equation}
which averages over pairs of galaxies with separation $\theta=|\mathbf{x}_i-\mathbf{x}_j|$, is expected to be sensitive to an ellipticity-dependent weighting scheme. Relevant for the estimation of cosmological parameters is the amplitude and shape of the shear correlation function $\xi_g(\theta)$. A straightforward calculation (outlined in \autoref{sec:biased_C}) shows that the normalized correlation function is given by
\begin{equation}
\label{eq:biased_C}
\frac{\langle w_i\bepsilon_i w_j\bepsilon_j\rangle}{\langle w_i w_j\rangle} \approx
\frac{c^2}{w_0^2}\bigl[\sigma_e^2 + \sigma_g^2\bigl]^2 +\, \xi_g(\theta)+ \frac{c^2}{w_0^2}\xi_g^2(\theta)
\end{equation}
where $\sigma_g^2$ denotes the variance of the shear field. Since the constant terms in the equation above can be determined by looking at separations $\theta$ where we expect the cosmological lensing signal to vanish, the correlation function is, surprisingly, largely unaffected by the weighting scheme. Only at very small scales, it is steepened due to the last term in \autoref{eq:biased_C}, but we expect this effect to be clearly subdominant compared to the influence of baryonic physics at these small scales.

In contrast to the outlier rejection bias -- which corresponds to a binary weight: either 1 or 0 -- this bias is less severe for large ellipticities, but it generally affects all galaxies, including those with smaller ellipticities, for which the creation of outliers is not a significant issue. Consequently, whenever weighted ellipticities are inserted into statistics that have been derived for unweighted ellipticities, correction terms (such as \autoref{eq:weighting_bias}) need to be applied, which requires accurate knowledge of the ellipticity-dependence of the weighting scheme.

\section{Summary, conclusions, and outlook}
\label{sec:discussion}

We compiled theoretical and practical evidence that pixel noise biases shear estimates at three subsequent levels: 1) the propagation of the pixel noise into the non-linear quantity ellipticity; 2) additional uncertainties and biases introduced by ellipticity-measurement methods; 3) the application of statistics to infer the gravitational shear from a sample of ellipticity measurements,  which are either unaware of the presence of pixel noise or themselves non-linear and thus biased. 

We summarize our findings in \autoref{tab:summary}. In practice, lensing analyses are affected by a combination of the aforementioned biases, most of which are negative. Considering these findings, it is not surprising that every investigation of any shear estimation methodology the authors are aware of shows a weakening response to shear with increasing noise level. It is inevitable. Even though the details differ, the biases we revealed or reinvestigated generally not only depend on the object significance but also on its ellipticity. We therefore recommend to extend shear accuracy test programs to also inspect trends with ellipticity.

That biases occur at several stages of the analysis pipeline leads to a unfortunate interdependence of the idiosyncrasies of the image data (foremost the galaxy ellipticity and signal-to-noise distributions), shape-measurement method, and shear statistic. This means that a setup which has been found to work well in one situation does not necessarily perform so well in others. Given the scope and fundamental nature of the majority of these biases, we do not believe that the needs of upcoming lensing surveys in terms of accuracy and reliability can be met without a substantial effort in correcting for or avoiding biased ellipticity and shear estimates.

Method-dependent biases can be studied with simulated images. Special attention should be paid to cases where the measured ellipticity distribution deviates from the expected Marsaglia distribution. For instance, a pile-up of samples shortly before the unit circle is a clear indication of a bias introduced by the shape measurement code in order to prevent unphysical outliers.

To correct biases in an actual measurement, one needs to be able to identify those parameters that determine the bias. As we argued, this is mainly the significance and the ellipticity, but other factors, such as the radial profile or changes of the ellipticity with radius \citep{Bernstein10.1}, will also play a role. Therefore, correction schemes need to fully consider the performance of the shape measurement method as a function of all relevant parameters as well as the fact that these parameters themselves can only be inferred with a certain precision from the image data \citep{Kacprzak12.1}. This is computationally and practically challenging.

Instead of correcting ellipticity estimates, which also does not completely eliminate the problem of noise for the shear statistic, we should seek solutions that avoid the biases, for instance by staying linear in the data for as long as possible \citep[exemplified by stacking methods in][]{Bridle10.1}. We conclude by sketching out an alternative idea. Since we worked out the theoretical form of the noisy ellipticity distribution, we are able to predict the (biased) outcome of a measurement given an assumed galaxy ellipticity and noise level. As we showed in \autoref{sec:measurements}, we can also model the errors shape measurement methods exhibit once we assume to know the underlying galaxy parameters. We can thus ask the question: How likely is a certain ellipticity given a measured one and its measurement errors. The result will be an ellipticity likelihood for each galaxy that incorporates the non-linearities of the measurement process. Combining full likelihoods should finally lead to unbiased shear estimates. Whether this idea works in practice remains to be seen.

To facilitate the review of our findings and to support forthcoming development, we make the code used in this work public. The {\sc python} implementation comprises the computation of the Marsaglia distribution with efficient sample generation as well as all shear estimators of \autoref{sec:statistics} and is available at \href{https://github.com/pmelchior/epsnoise}{this URL}.

\section*{Acknowledgments}
PM is supported by the U.S. Department of Energy under Contract No. DE- FG02-91ER40690. MV is supported by a STFC Rolling Grant RA0888. The authors want to thank Alan Heavens for pointing out the origin of $\rho_n=\tfrac{1}{3}$ in \autoref{eq:rho_13}.

{\small
\bibliography{../references}

\begin{thebibliography}{39}
\expandafter\ifx\csname natexlab\endcsname\relax\def\natexlab#1{#1}\fi

\bibitem[{{Amara} \& {Refregier}(2008)}]{Amara08.1}
\href{http://adsabs.harvard.edu/abs/2008MNRAS.391..228A}{{Amara}, A., \&
  {Refregier}, A. 2008, \mnras, 391, 228}

\bibitem[{{Andrae} {et~al.}(2011){Andrae}, {Melchior}, \&
  {Jahnke}}]{Andrae11.2}
\href{http://adsabs.harvard.edu/abs/2011MNRAS.417.2465A}{{Andrae}, R.,
  {Melchior}, P., \& {Jahnke}, K. 2011, \mnras, 417, 2465}

\bibitem[{{Bartelmann} \& {Schneider}(2001)}]{Bartelmann01.1}
\href{http://adsabs.harvard.edu/abs/2001PhR...340..291B}{{Bartelmann}, M.,
  \& {Schneider}, P. 2001, \physrep, 340, 291}

\bibitem[{{Bernstein}(2010)}]{Bernstein10.1}
\href{http://adsabs.harvard.edu/abs/2010MNRAS.406.2793B}{{Bernstein}, G.~M.
  2010, \mnras, 406, 2793}

\bibitem[{{Bernstein} \& {Jarvis}(2002)}]{Bernstein02.1}
\href{http://adsabs.harvard.edu/abs/2002AJ....123..583B}{{Bernstein}, G.~M.,
  \& {Jarvis}, M. 2002, \aj, 123, 583}

\bibitem[{{Bridle} {et~al.}(2010){Bridle}, {Balan}, {Bethge}, {Gentile},
  {Harmeling}, {Heymans}, {Hirsch}, {Hosseini}, {Jarvis}, {Kirk}, {Kitching},
  {Kuijken}, {Lewis}, {Paulin-Henriksson}, {Sch{\"o}lkopf}, {Velander},
  {Voigt}, {Witherick}, {Amara}, {Bernstein}, {Courbin}, {Gill}, {Heavens},
  {Mandelbaum}, {Massey}, {Moghaddam}, {Rassat}, {R{\'e}fr{\'e}gier}, {Rhodes},
  {Schrabback}, {Shawe-Taylor}, {Shmakova}, {van Waerbeke}, \&
  {Wittman}}]{Bridle10.1}
\href{http://adsabs.harvard.edu/abs/2010MNRAS.405.2044B}{{Bridle}, S.
  {et~al.} 2010, \mnras, 405, 2044}

\bibitem[{{Erben} {et~al.}(2001){Erben}, {Van Waerbeke}, {Bertin}, {Mellier},
  \& {Schneider}}]{Erben01.1}
\href{http://adsabs.harvard.edu/abs/2001A&A...366..717E}{{Erben}, T., {Van
  Waerbeke}, L., {Bertin}, E., {Mellier}, Y., \& {Schneider}, P. 2001, \aap,
  366, 717}

\bibitem[{{Gruen} {et~al.}(2010){Gruen}, {Seitz}, {Koppenhoefer}, \&
  {Riffeser}}]{Gruen10.1}
\href{http://adsabs.harvard.edu/abs/2010ApJ...720..639G}{{Gruen}, D.,
  {Seitz}, S., {Koppenhoefer}, J., \& {Riffeser}, A. 2010, \apj, 720, 639}

\bibitem[{{H{\"a}ussler} {et~al.}(2007){H{\"a}ussler}, {McIntosh}, {Barden},
  {Bell}, {Rix}, {Borch}, {Beckwith}, {Caldwell}, {Heymans}, {Jahnke}, {Jogee},
  {Koposov}, {Meisenheimer}, {S{\'a}nchez}, {Somerville}, {Wisotzki}, \&
  {Wolf}}]{Haeussler07.1}
\href{http://adsabs.harvard.edu/abs/2007ApJS..172..615H}{{H{\"a}ussler}, B.
  {et~al.} 2007, \apjs, 172, 615}

\bibitem[{{Heymans} {et~al.}(2006){Heymans}, {Van Waerbeke}, {Bacon}, {Berge},
  {Bernstein}, {Bertin}, {Bridle}, {Brown}, {Clowe}, {Dahle}, {Erben}, {Gray},
  {Hetterscheidt}, {Hoekstra}, {Hudelot}, {Jarvis}, {Kuijken}, {Margoniner},
  {Massey}, {Mellier}, {Nakajima}, {Refregier}, {Rhodes}, {Schrabback}, \&
  {Wittman}}]{Heymans06.1}
\href{http://adsabs.harvard.edu/abs/2006MNRAS.368.1323H}{{Heymans}, C.
  {et~al.} 2006, \mnras, 368, 1323}

\bibitem[{{Hirata} {et~al.}(2004){Hirata}, {Mandelbaum}, {Seljak}, {Guzik},
  {Padmanabhan}, {Blake}, {Brinkmann}, {Bud{\'a}vari}, {Connolly}, {Csabai},
  {Scranton}, \& {Szalay}}]{Hirata04.1}
\href{http://adsabs.harvard.edu/abs/2004MNRAS.353..529H}{{Hirata}, C.~M.
  {et~al.} 2004, \mnras, 353, 529}

\bibitem[{{Hoekstra} {et~al.}(2000){Hoekstra}, {Franx}, \&
  {Kuijken}}]{Hoekstra00.1}
\href{http://adsabs.harvard.edu/abs/2000ApJ...532...88H}{{Hoekstra}, H.,
  {Franx}, M., \& {Kuijken}, K. 2000, \apj, 532, 88}

\bibitem[{{Huterer} {et~al.}(2006){Huterer}, {Takada}, {Bernstein}, \&
  {Jain}}]{Huterer06.1}
\href{http://adsabs.harvard.edu/abs/2006MNRAS.366..101H}{{Huterer}, D.,
  {Takada}, M., {Bernstein}, G., \& {Jain}, B. 2006, \mnras, 366, 101}

\bibitem[{{Kacprzak} {et~al.}(2012){Kacprzak}, {Zuntz}, {Rowe}, {Bridle},
  {Refregier}, {Amara}, {Voigt}, \& {Hirsch}}]{Kacprzak12.1}
\href{http://adsabs.harvard.edu/abs/2012arXiv1203.5049K}{{Kacprzak}, T.,
  {Zuntz}, J., {Rowe}, B., {Bridle}, S., {Refregier}, A., {Amara}, A., {Voigt},
  L., \& {Hirsch}, M. 2012, ArXiv e-prints, arXiv:1203.5049}

\bibitem[{{Kaiser} {et~al.}(1995){Kaiser}, {Squires}, \&
  {Broadhurst}}]{Kaiser95.1}
\href{http://adsabs.harvard.edu/abs/1995ApJ...449..460K}{{Kaiser}, N.,
  {Squires}, G., \& {Broadhurst}, T. 1995, \apj, 449, 460}

\bibitem[{{Kitching} {et~al.}(2012){Kitching}, {Balan}, {Bridle}, {Cantale},
  {Courbin}, {Gentile}, {Gill}, {Harmeling}, {Heymans}, {Hirsch}, {Kacprzak},
  {Kirkby}, {Margala}, {Massey}, {Melchior}, {Nurbaeva}, {Patton}, {Rhodes},
  {Rowe}, {Taylor}, {Tewes}, {Viola}, {Witherick}, {Voigt}, {Young}, \&
  {Zuntz}}]{Kitching12.1}
\href{http://adsabs.harvard.edu/abs/2012arXiv1202.5254K}{{Kitching}, T.~D.
  {et~al.} 2012, ArXiv e-prints, arXiv:1202.5254}

\bibitem[{{Kitching} {et~al.}(2008){Kitching}, {Miller}, {Heymans}, {van
  Waerbeke}, \& {Heavens}}]{Kitching08.1}
\href{http://adsabs.harvard.edu/abs/2008MNRAS.390..149K}{{Kitching}, T.~D.,
  {Miller}, L., {Heymans}, C.~E., {van Waerbeke}, L., \& {Heavens}, A.~F. 2008,
  \mnras, 390, 149}

\bibitem[{{Lewis}(2009)}]{Lewis09.1}
\href{http://adsabs.harvard.edu/abs/2009MNRAS.398..471L}{{Lewis}, A. 2009,
  \mnras, 398, 471}

\bibitem[{{Mandelbaum} {et~al.}(2012){Mandelbaum}, {Hirata}, {Leauthaud},
  {Massey}, \& {Rhodes}}]{Mandelbaum12.1}
\href{http://adsabs.harvard.edu/abs/2012MNRAS.420.1518M}{{Mandelbaum}, R.,
  {Hirata}, C.~M., {Leauthaud}, A., {Massey}, R.~J., \& {Rhodes}, J. 2012,
  \mnras, 420, 1518}

\bibitem[{Marsaglia(1965)}]{Marsaglia65.1}
\href{http://www.jstor.org/stable/2283145}{Marsaglia, G. 1965, Journal of the American Statistical Association, 60, pp.
  193}

\bibitem[{Marsaglia(2006)}]{Marsaglia06.1}
\href{http://www.jstatsoft.org/v16/i04}{Marsaglia, G. 2006, Journal of Statistical Software, 16, 1}

\bibitem[{{Massey} {et~al.}(2007){Massey}, {Heymans}, {Berg{\'e}}, {Bernstein},
  {Bridle}, {Clowe}, {Dahle}, {Ellis}, {Erben}, {Hetterscheidt}, {High},
  {Hirata}, {Hoekstra}, {Hudelot}, {Jarvis}, {Johnston}, {Kuijken},
  {Margoniner}, {Mandelbaum}, {Mellier}, {Nakajima}, {Paulin-Henriksson},
  {Peeples}, {Roat}, {Refregier}, {Rhodes}, {Schrabback}, {Schirmer}, {Seljak},
  {Semboloni}, \& {van Waerbeke}}]{Massey07.1}
\href{http://adsabs.harvard.edu/abs/2007MNRAS.376...13M}{{Massey}, R.
  {et~al.} 2007, \mnras, 376, 13}

\bibitem[{Massey {et~al.}(2004)Massey, Refregier, Conselice, \&
  Bacon}]{Massey04.1}
\href{http://adsabs.harvard.edu/abs/2004MNRAS.348..214M}{Massey, R.,
  Refregier, A., Conselice, C.~J., \& Bacon, D.~J. 2004, \mnras, 348, 214}

\bibitem[{{Melchior} {et~al.}(2010){Melchior}, {B{\"o}hnert}, {Lombardi}, \&
  {Bartelmann}}]{Melchior10.1}
\href{http://adsabs.harvard.edu/abs/2010A&A...510A..75M}{{Melchior}, P.,
  {B{\"o}hnert}, A., {Lombardi}, M., \& {Bartelmann}, M. 2010, \aap, 510, A75+}

\bibitem[{{Melchior} {et~al.}(2011){Melchior}, {Viola}, {Sch{\"a}fer}, \&
  {Bartelmann}}]{Melchior11.1}
\href{http://adsabs.harvard.edu/abs/2011MNRAS.412.1552M}{{Melchior}, P.,
  {Viola}, M., {Sch{\"a}fer}, B.~M., \& {Bartelmann}, M. 2011, \mnras, 412,
  1552}

\bibitem[{{Meneghetti} {et~al.}(2008){Meneghetti}, {Melchior}, {Grazian}, {De
  Lucia}, {Dolag}, {Bartelmann}, {Heymans}, {Moscardini}, \&
  {Radovich}}]{Meneghetti08.1}
\href{http://adsabs.harvard.edu/abs/2008A&A...482..403M}{{Meneghetti}, M.
  {et~al.} 2008, \aap, 482, 403}

\bibitem[{{Miller} {et~al.}(2007){Miller}, {Kitching}, {Heymans}, {Heavens}, \&
  {van Waerbeke}}]{Miller07.1}
\href{http://adsabs.harvard.edu/abs/2007MNRAS.382..315M}{{Miller}, L.,
  {Kitching}, T.~D., {Heymans}, C., {Heavens}, A.~F., \& {van Waerbeke}, L.
  2007, \mnras, 382, 315}

\bibitem[{{Moffat}(1969)}]{Moffat69.1}
\href{http://adsabs.harvard.edu/abs/1969A&A.....3..455M}{{Moffat}, A.~F.~J.
  1969, \aap, 3, 455}

\bibitem[{{Ngan} {et~al.}(2009){Ngan}, {van Waerbeke}, {Mahdavi}, {Heymans}, \&
  {Hoekstra}}]{Ngan09.1}
\href{http://adsabs.harvard.edu/abs/2009MNRAS.396.1211N}{{Ngan}, W., {van
  Waerbeke}, L., {Mahdavi}, A., {Heymans}, C., \& {Hoekstra}, H. 2009, \mnras,
  396, 1211}

\bibitem[{{Okura} \& {Futamase}(2009)}]{Okura09.1}
\href{http://adsabs.harvard.edu/abs/2009ApJ...699..143O}{{Okura}, Y., \&
  {Futamase}, T. 2009, \apj, 699, 143}

\bibitem[{{Refregier}(2003)}]{Refregier03.1}
\href{http://adsabs.harvard.edu/abs/2003MNRAS.338...35R}{{Refregier}, A.
  2003, \mnras, 338, 35}

\bibitem[{{Refregier} {et~al.}(2012){Refregier}, {Kacprzak}, {Amara}, {Bridle},
  \& {Rowe}}]{Refregier12.1}
\href{http://adsabs.harvard.edu/abs/2012arXiv1203.5050R}{{Refregier}, A.,
  {Kacprzak}, T., {Amara}, A., {Bridle}, S., \& {Rowe}, B. 2012, ArXiv
  e-prints, arXiv:1203.5050}

\bibitem[{{Schneider} \& {Seitz}(1995)}]{Schneider95.1}
\href{http://adsabs.harvard.edu/abs/1995A&A...294..411S}{{Schneider}, P., \&
  {Seitz}, C. 1995, \aap, 294, 411}

\bibitem[{{Seitz} \& {Schneider}(1997)}]{Seitz97.1}
\href{http://adsabs.harvard.edu/abs/1997A&A...318..687S}{{Seitz}, C., \&
  {Schneider}, P. 1997, \aap, 318, 687}

\bibitem[{{Sersic}(1968)}]{Sersic68.1}
\href{http://adsabs.harvard.edu/abs/1968adga.book.....S}{{Sersic}, J.~L.
  1968, Atlas de galaxias australes (Cordoba, Argentina: Observatorio
  Astronomico, 1968)}

\bibitem[{{Tewes} {et~al.}(2012){Tewes}, {Cantale}, {Courbin}, {Kitching}, \&
  {Meylan}}]{Tewes12.1}
\href{http://adsabs.harvard.edu/abs/2012arXiv1203.4429T}{{Tewes}, M.,
  {Cantale}, N., {Courbin}, F., {Kitching}, T.~D., \& {Meylan}, G. 2012, ArXiv
  e-prints, arXiv:1203.4429}

\bibitem[{{Viola} {et~al.}(2011){Viola}, {Melchior}, \&
  {Bartelmann}}]{Viola11.1}
\href{http://adsabs.harvard.edu/abs/2011MNRAS.410.2156V}{{Viola}, M.,
  {Melchior}, P., \& {Bartelmann}, M. 2011, \mnras, 410, 2156}

\bibitem[{{Voigt} \& {Bridle}(2010)}]{Voigt10.1}
\href{http://adsabs.harvard.edu/abs/2010MNRAS.404..458V}{{Voigt}, L.~M., \&
  {Bridle}, S.~L. 2010, \mnras, 404, 458}

\bibitem[{{Weinberg} {et~al.}(2012){Weinberg}, {Mortonson}, {Eisenstein},
  {Hirata}, {Riess}, \& {Rozo}}]{Weinberg12.1}
\href{http://adsabs.harvard.edu/abs/2012arXiv1201.2434W}{{Weinberg}, D.~H.,
  {Mortonson}, M.~J., {Eisenstein}, D.~J., {Hirata}, C., {Riess}, A.~G., \&
  {Rozo}, E. 2012, ArXiv e-prints, arXiv:1201.2434}

\end{thebibliography}
}

\appendix

\section{Moment-based ellipticity definitions}
\label{sec:eps_chi}

\begin{figure}
\includegraphics[width=\linewidth]{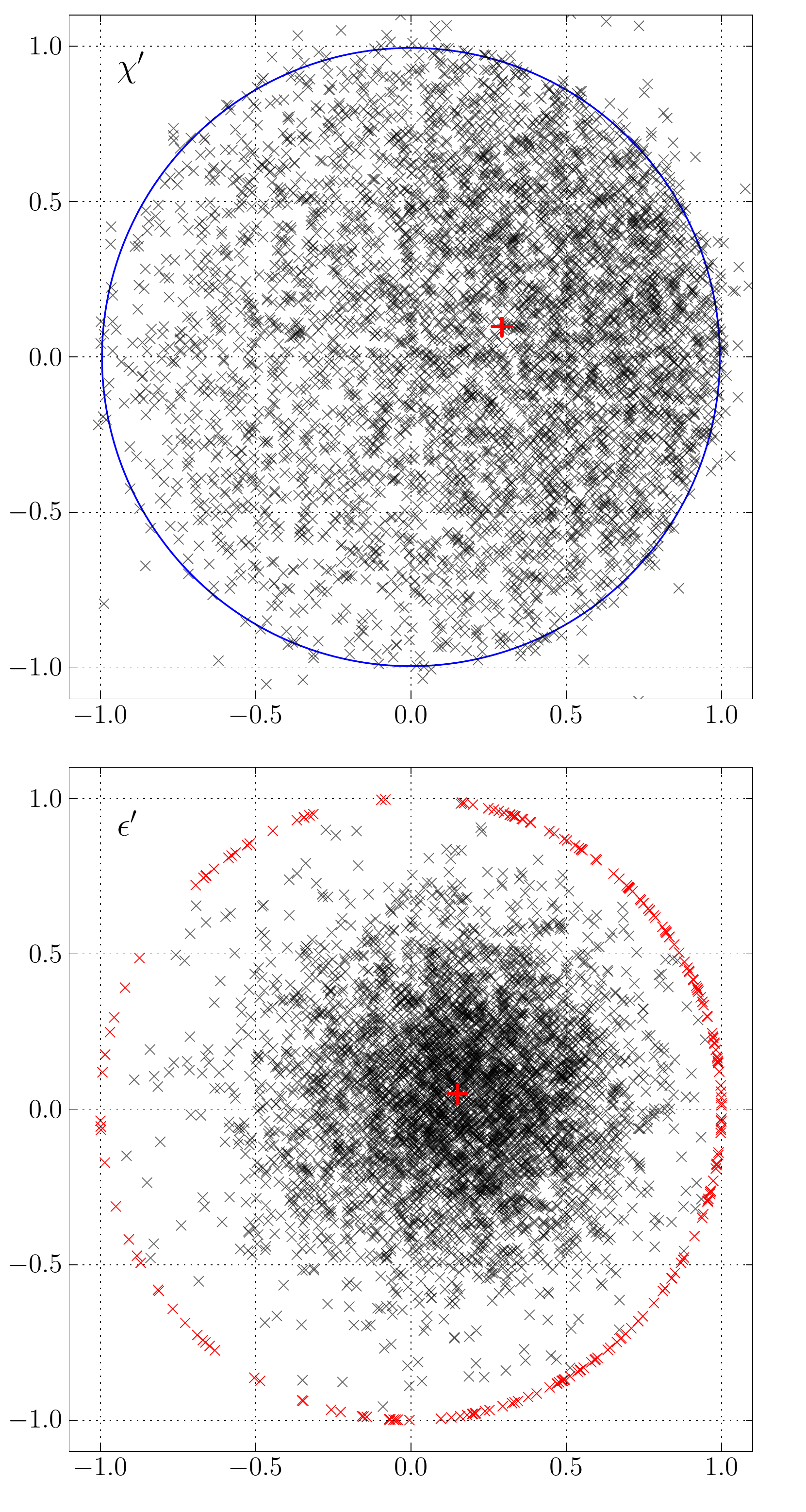}
\caption{Ellipticity distributions of measured $\bchi^\prime$ (\emph{top}) and measured $\bepsilon^\prime$ (\emph{bottom}) from a synthetic catalog of 5,000 objects with intrinsic dispersion $\sigma_e=0.3$, constant shear $\mathbf{g} = (0.15,\,0.05)$, indicated by red cross markers, and pixel noise equivalent to $\nu=15$. The circle in the top plot corresponds to the limit of intrinsic $|\bepsilon|<1$, the red points in the bottom plot indicate galaxies with $|\bepsilon^\prime|\geq1$. The distributions are created with the algorithm outlined in \autoref{sec:sampling}.}
\label{fig:eps_planes}
\end{figure}

In \autoref{eq:ellipticities} we have introduced a two particular forms of the complex ellipticity, which are related according to \citep{Bartelmann01.1}
\begin{equation}
\label{eq:eps2chi}
\bchi=\frac{2\bepsilon}{1+|\bepsilon|^2} \ \ \text{or}\ \ \bepsilon = \frac{\bchi}{1+\sqrt{1-|\bchi|^2}}.
\end{equation}
Since the only significant difference between the two definitions occurs in the denominator, they share the same complex phase, but have different amplitude. There are two important distinctions between these definitions. First, the relation between gravitational shear $\mathbf{g}$ and mean ellipticity is
\begin{equation}
\label{eq:mean_e}
\mathbf{g} = \langle\bepsilon\rangle \ \ \mathrm{or} \ \ \mathbf{g} = \frac{\langle\bchi\rangle}{2+\sigma_\chi^2} + \mathcal{O}(\langle\bchi^3\rangle),  
\end{equation}
where the averages are taken over galaxies affected by a constant shear $\mathbf{g}$. That means, $\bepsilon$ is an unbiased estimator of the shear \citep{Seitz97.1}, while $\bchi$ needs a so-called \emph{responsivity} correction (the denominator in the equation above), and even then it  remains an approximate estimator of $\mathbf{g}$ (see \citet{Viola11.1} for a discussion of higher-order corrections to this relation). 

The reason for the wide-spread use of $\bchi$ as shear estimator rather than $\bepsilon$ stems from their second distinction, the distribution under pixel noise. If we assume the pixel noise to be uncorrelated and Gaussian, the second-order moments share this property due to their linearity in the data. However, both $\bepsilon$ and $\bchi$ are ratios of combinations of second-order moments. In the case of $\bchi$, numerator and denominator are linear combination of moments and thus still follow Gaussian distributions such that their ratio follows the Marsaglia distribution of \autoref{eq:marsaglia}, a generalization of the Cauchy distribution (\autoref{eq:cauchy}). The Cauchy distribution is known for its diverging variance, which allows arbitrarily large errors with finite probability, namely when $Q_{11}+Q_{22}\rightarrow 0$. In contrast, due to the non-linear term $\sqrt{Q_{11}Q_{22}-Q_{12}^2}$ in the denominator of \autoref{eq:ellipticity}, $\bepsilon$ has a much more complicated distribution. Instead of allowing infinite errors, the additional term in the denominator can lead to a complex phase, which alters the orientation of $\bepsilon$ in the cases where $\bchi$ would come to lie outside of the unit circle of viable ellipticities. This effectively couples the two components of $\bepsilon$ and leads to two separate distributions, the ordinary one within the unit circle and the one on the unit circle, which renders a theoretical description much more difficult.  

We show the different distributions of $\bepsilon$ and $\bchi$ under noise in \autoref{fig:eps_planes}.
While $\bchi^\prime$ shows a continuous distribution across the edge of the unit circle, outlier ellipticities (red points) are located right on the unit circle in $\bepsilon^\prime$-space. Performing a projection onto one component or computing the absolute value of $\bepsilon^\prime$ would thus lead to an enhanced probability of measurements close to the unit circle.
Moreover, global statistics such as the mean or the variance of the $\bepsilon^\prime$ distribution are significantly altered by the presence of the ring at $|\bepsilon^\prime|=1$.

In summary, while $\bepsilon$ is theoretically an unbiased estimator of the shear, its distribution under noise effectively undermines this property. On the other hand, the distribution of $\bchi$ is much easier to describe and statistics thereof are fairly robust against noise, but the interpretation of these statistics is hampered by the non-linear relation to shear. One therefore needs to choose the ellipticity estimator based on the application at hand, depending on what kind of drawback can most effectively be dealt with.

\section{Sampling the noisy ellipticity distribution}
\label{sec:sampling}

We present the recipe to simulate a realistic ellipticity distribution, considering the effects of non-linear error propagation discussed in \autoref{sec:error_propagation}. Instead of sampling from the complicated Marsaglia distribution of \autoref{eq:marsaglia}, we follow the path of the pixel noise, i.e. we compute the means and errors for the moment combinations $w=Q_{11} + Q_{22}$ and $z=Q_{11} + Q_{22}$ and then their ratio $\chi=\tfrac{z}{w}$. 

For a fast and analytic moment calculation, we assume an elliptical Gaussian shape for both galaxy and/or weight function and rotate into a frame, such that its semi-major axis is aligned with the 1-direction,
\begin{equation}
W(\mathbf{x}) = \exp\Big[\frac{-(1-\epsilon)^2 x_1^2 - (1+\epsilon)^2 x_2^2}{2s^2}\Big].
\end{equation}
The size of the galaxy is then defined by $s$ and its flux $F=\int d^2x W(\mathbf{x})$. Errors of the moments, measured with weight function $W$, are given by \citep{Melchior11.1}
\begin{equation}
\sigma_{i,j}^2 = \sigma_n^2\int d^2 x\ W^2(\mathbf{x}) x_1^{2i} x_2^{2j},
\end{equation}
where $\sigma_n^2$ denotes the pixel noise variance and $i,j$ describe a moment $\lbrace W\rbrace_{i,j} = \int d^2x\ W(x) x_1^i x_2^j$. Note that this notation differs from the notation used throughout the rest of this work. In the case of a Gaussian-shaped $W$, these errors can be analytically evaluated. The algorithm can trivially be extended to work on moments from arbitrarily shaped galaxies, measured from noise-free images. With the convolution relation in moment-space \citep[Equation 9 therein]{Melchior11.1}, also the effect of the PSF convolution can be taken into account.

\begin{compactenum}[1)]
\item For a source with ellipticity $\bepsilon$, we compute the modulus $\epsilon=|\bepsilon|$, its equivalent $\chi$ from \autoref{eq:eps2chi}, and the phase $\phi = \arg(\bepsilon)$.
\item For a Gaussian shape, the scale $s$ defines the sum of second moments,  $w=F s^2$. Then, $z=\chi w$.
\item We adopt the definition of significance from \citet[Equation 16 therein]{Erben01.1} and insert the error of the flux $F$ (the $\lbrace W\rbrace_{0,0}$ moment in the notation used above):  $\nu = \frac{F}{\sigma_n\sqrt{\mathrm{\pi}}s}\sqrt{(1+\epsilon)(1-\epsilon)}$. Thus, if we specify $\nu$ we get the pixel noise level $\sigma_n$.
\item The Gaussian errors $\sigma_{ij}$ for the second moments $Q_{ij}$ are then given by
\begin{equation}
\begin{split}
&\sigma_{11}^2 = \sigma_{2,0}^2 = \sigma_n^2 \frac{3\mathrm{\pi}}{4}\frac{s^6}{(1-\epsilon)^5(1+\epsilon)}\\
&\sigma_{12}^2 = \sigma_{1,1}^2 = \sigma_n^2 \frac{\mathrm{\pi}}{4}\frac{s^6}{(1-\epsilon)^3(1+\epsilon)^3}\\
&\sigma_{22}^2 = \sigma_{0,2}^2 = \sigma_n^2 \frac{3\mathrm{\pi}}{4}\frac{s^6}{(1-\epsilon)(1+\epsilon)^5}
\end{split}
\end{equation}
\item For the errors  $\Delta Q_{11}$ and $\Delta Q_{22}$, we sample from a correlated bi-variate Gaussian, whose  covariance matrix is given by
\begin{equation}
\mathsf{S}_{11,22}=\begin{pmatrix} \sigma_{11}^2 & \rho_n\sigma_{11}\sigma_{22}
\\ \rho_n\sigma_{11}\sigma_{22} & \sigma_{22}^2.
\end{pmatrix}
\end{equation}
(cf. discussion that led to \autoref{eq:Swz}). This can be realized by applying the Cholesky decomposition
\begin{equation}
\mathsf{S}_{11,22} = \mathsf{A}\mathsf{A^T} \rightarrow \mathsf{A} = \begin{pmatrix} \sigma_{11} & 0\\ \rho_n\sigma_{22} & \sigma_{22} \sqrt{1-\rho_n^2}
\end{pmatrix}
\end{equation}
to a vector of two $\mathcal{N}(0,1)$ variates.
\item Then, noisy samples $w^\prime = w + \Delta Q_{11} + \Delta Q_{22}$ and $z^\prime = z + \Delta Q_{11} - \Delta Q_{22}$ exhibit the correct variances and correlation given by \autoref{eq:Swz}, and their ratio $\chi_1^\prime = \tfrac{z^\prime}{w^\prime}$ is distributed according to the Marsaglia distribution of \autoref{eq:marsaglia}.
\item The second component of $\bchi$, has an independent numerator $\Delta Q_{12} \sim\mathcal{N}(0,\sigma_{12}^2)$, but the same denominator: $\chi_2^\prime = \tfrac{2\Delta Q_{12}}{w^\prime}$.
\item The original orientation is recovered by $\bchi^\prime \rightarrow \bchi^\prime \mathrm{e}^{\mathrm{i}\phi}$. Then, $\bepsilon^\prime$ can be obtained from \autoref{eq:eps2chi}.
\end{compactenum}
Since the only terms entering the Marsaglia distribution are ratios of moments and errors, and both scale as $s^2$, we can set $s=1$. Similarly, the significance $\nu$ depends on the ratio of  flux $F$ and noise dispersion $\sigma_n$, so that we can also set $F=1$ without changing the results. Effectively, one only has to specify the ellipticity $\bepsilon$ and the significance $\nu$ to uniquely describe the effects of noise on the ellipticity. All information about the galactic shape, which determines the moments of the brightness distribution and their errors, is then internally computed (assuming a Gaussian radial profile).

Realistic distributions, such as the ones shown in \autoref{fig:eps_planes}, additionally need to start from a decent intrinsic distributions of ellipticities $\bepsilon^s$. With the common assumption of each ellipticity component being drawn from an independent Gaussian distribution of some dispersion $\sigma_e\approx 0.3$, their modulus is drawn from the Rayleigh distribution $f(\epsilon) = \frac{\epsilon}{\sigma_e^2} \mathrm{e}^{-\epsilon^2/2\sigma_e^2}$ (cf. \autoref{fig:gems_eps} for an observed ellipticity distribution), and the orientation is uniform, $\mathcal{U}(0,\mathrm{\pi})$. Finally, a sheared ellipticity distribution can be obtained from \citep{Seitz97.1}
\begin{equation}
\label{eq:eps_shear}
\bepsilon = \frac{\bepsilon^s+\mathbf{g}}{1+\bepsilon^s\mathbf{g}^*}.
\end{equation}

We implemented the entire sampling procedure (as well as the theoretical form of the Marsaglia distribution and the shear estimators described in \autoref{sec:statistics}) in {\sc python}. The code is open-source and available at \href{https://github.com/pmelchior/epsnoise}{https://github.com/pmelchior/epsnoise}.

\section{Weighting-scheme induced bias} 
\label{sec:weighting_bias}
We seek an approximate analytical description of the bias an ellipticity-dependent weighting scheme introduces on globally acting statistics such as the average and the two-point correlation function. Therefore, we choose a coordinate frame, which is aligned with the semi-major axis of each galaxy, such that we only have to consider the one component of $\bepsilon$. We start by Taylor-expanding the ellipticity dependence of the weighting scheme to first order,
\begin{equation}
w(\epsilon^\prime) = w_0 + \frac{\partial w(\epsilon)}{\partial \epsilon}\Big|_{\epsilon^\prime} \ \epsilon^\prime + \mathcal{O}(\epsilon^{\prime 2}).
\end{equation}
Assuming the most simple form of a positive offset $w_0>0$ and small, constant slope $\partial w(|\epsilon|)/\partial |\epsilon|=c$, we can simplify the previous equation to
\begin{equation}
\label{eq:weighting_scheme}
w(\epsilon) = w_0 
\begin{cases}
+\ c\epsilon^\prime & \ \ \text{if}\ \ \epsilon > 0\\
-\ c\epsilon^\prime & \ \ \text{if}\ \  \epsilon < 0.\\
\end{cases}
\end{equation}
Now we split the integrals in of \autoref{eq:weighted_av} into the lower and upper half (which allows us to apply the reduced weights also to negative $\epsilon$):
\begin{equation}
\begin{split}
\int d\epsilon\ w(\epsilon) \epsilon\, p(\epsilon) &=\ w_0\int\limits_{-\infty}^0 d\epsilon\ \epsilon\, p(\epsilon) - c\int\limits_{-\infty}^0 d\epsilon\ \epsilon^2 p(\epsilon)\\ 
&+\ w_0\int\limits_0^{\infty} d\epsilon\ \epsilon\, p(\epsilon) + c\int\limits_0^{\infty} d\epsilon\ \epsilon^2 p(\epsilon)
\end{split}
\end{equation}
and likewise for the denominator. Since there is no sign-flip in the terms with $w_0$, we can combine these integrals again and exploit $\int d\epsilon\, p(\epsilon) = 1$ and $\int d\epsilon\, \epsilon\, p(\epsilon) = g$. For the $c$-terms, we need to carry out the integration over the ellipticity distribution explicitly, simply because the presence of the shear shifts and skews the distribution such that the $c$-terms in the equation above do not exactly cancel. We therefore assume the pre-lensing distribution to be Gaussian with dispersion $\sigma_e$ and the shear to only shift the entire distribution without changing its shape: $p(\epsilon)\rightarrow \mathcal{N}(\epsilon-g,\sigma_e)$. This assumption will restrict our derivation to small shears. We thus linearize the resulting integrals to first order in the shear and obtain
\begin{equation}
\begin{split}
&\int\limits_{-\infty}^0 d\epsilon\ \epsilon\, \mathcal{N}(\epsilon-g,\sigma_e) - \int\limits_0^{\infty} d\epsilon\ \epsilon\, \mathcal{N}(\epsilon-g,\sigma_e) \approx -\sqrt{\frac{2}{\pi}}\sigma_e\\
&\int\limits_{-\infty}^0 d\epsilon\ \epsilon^2 \mathcal{N}(\epsilon-g,\sigma_e) - \int\limits_0^{\infty} d\epsilon\ \epsilon^2 \mathcal{N}(\epsilon-g,\sigma_e) \approx -2\sqrt{\frac{2}{\pi}}\sigma_e\ g.
\end{split}
\end{equation}
Inserting all terms into \autoref{eq:weighted_av} yields \autoref{eq:weighting_bias}. The result is accurate to first order in $w(\epsilon)$, $g$, and $\tfrac{c}{w_0}$.

\subsection{Shear correlation function}
\label{sec:biased_C}
In the limit of small shear, we  can simplify \autoref{eq:eps_shear} between observed and pre-lensing ellipticities, $\bepsilon\rightarrow\bepsilon^s + \mathbf{g}$,
which corresponds to the case above: the shear only shifts the ellipticity distribution, but does not skew it. When considering the weighting scheme from \autoref{eq:weighting_scheme}, we get the correlations functions of the weights
\begin{equation}
\label{eq:ww}
\langle w_i w_j\rangle(\theta) = \langle (w_0 + c\bepsilon_i)(w_0 + c\bepsilon_j)\rangle = w_0^2 + c^2\langle\bepsilon_i\bepsilon_j\rangle(\theta)
\end{equation}
and of the weighted ellipticities
\begin{equation}
\label{eq:wewe}
\langle w_i\bepsilon_i w_j\bepsilon_j\rangle(\theta) = w_0^2\langle\bepsilon_i\bepsilon_j\rangle(\theta) + c^2\langle\bepsilon_i^2\bepsilon_j^2\rangle(\theta),
\end{equation}
where $\langle\bepsilon_i\bepsilon_j\rangle$ is given by \autoref{eq:epseps} and
\begin{equation}
\langle\bepsilon_i^2\bepsilon_j^2\rangle(\theta) = \sigma_e^4 + 2\sigma_e^2 \sigma_g^2 + \sigma_g^4 + 2\xi_g^2(\theta)
\end{equation}
with $\sigma_g^2\equiv\langle \mathbf{g}_i^2\rangle$ being the variance of the shear field. In this derivation we assumed the intrinsic ellipticity field to be uncorrelated for separation $\theta>0$ and the shear field to be uncorrelated with the intrinsic ellipticity field, such that terms like $\langle\bepsilon_i \mathbf{g}_i\rangle$ and $\langle\bepsilon_i \mathbf{g}_j\rangle$ vanish. Expanding the ratio of \autoref{eq:ww} and \autoref{eq:wewe} to first order in $\tfrac{c^2}{w_0^2}$ yields \autoref{eq:biased_C}.
\label{lastpage}
\end{document}